\RequirePackage{lineno}
\documentclass[twocolumn,aps,tightenlinefs,superscriptaddress,preprintnumbers]{revtex4}
\usepackage{graphicx,xcolor,color,dcolumn,booktabs,bm,amsmath,amssymb}
\usepackage{hyperref}  
\usepackage[utf8]{inputenc} 
\usepackage[english]{babel} 
\usepackage{physics}
\usepackage{lineno}
\usepackage{orcidlink}

\input{belle2-symbols}

\begin{document}
	\hyphenpenalty=10000

	\title{\quad\\[0.1cm]\boldmath   Measurements of the branching fractions of $\Lambda_c^+\to\Lambda K^+ \pi^0$ and $\Lambda_c^+\to\Lambda K^+ \eta$ decays at Belle and Belle~II }

  \author{M.~Abumusabh\,\orcidlink{0009-0004-1031-5425}} 
  \author{I.~Adachi\,\orcidlink{0000-0003-2287-0173}} 
  \author{A.~Aggarwal\,\orcidlink{0000-0002-5623-3896}} 
  \author{Y.~Ahn\,\orcidlink{0000-0001-6820-0576}} 
  \author{H.~Aihara\,\orcidlink{0000-0002-1907-5964}} 
  \author{M.~Akdag\,\orcidlink{0009-0004-3728-1077}} 
  \author{N.~Akopov\,\orcidlink{0000-0002-4425-2096}} 
  \author{A.~Akram\,\orcidlink{0000-0003-0198-5852}} 
  \author{S.~Alghamdi\,\orcidlink{0000-0001-7609-112X}} 
  \author{M.~Alhakami\,\orcidlink{0000-0002-2234-8628}} 
% ...
\collaboration{The Belle and Belle II Collaborations}

	\begin{abstract}

We present a study of the three-body, singly Cabibbo-suppressed decays $\Lambda_c^+ \to \Lambda K^+ \pi^0$ and $\Lambda_c^+ \to \Lambda K^+ \eta$ using data recorded with the Belle and Belle~II detectors at the KEKB and SuperKEKB $e^+ e^-$ colliders. The data samples correspond to integrated luminosities of 980\invfb and 570\invfb, respectively. 
We measure their branching-fraction ratios relative to the Cabibbo-favored decay $\Lambda_c^+ \to \Lambda \pi^+ \pi^0$ as $\mathcal{B}(\Lambda_c^+ \to \Lambda K^+ \pi^0)/\mathcal{B}(\Lambda_c^+ \to \Lambda \pi^+ \pi^0)=(1.82 \pm 0.07 \pm 0.07)\%$, with significantly improved precision compared to the previous measurement, and $\mathcal{B}(\Lambda_c^+ \to \Lambda K^+ \eta)/\mathcal{B}(\Lambda_c^+ \to \Lambda \pi^+ \pi^0)=(8.73 \pm 0.35 \pm 0.43)\times10^{-3}$, corresponding to the first observation of the $\Lambda_c^+ \to \Lambda K^+ \eta$ decay with a statistical significance exceeding 10 standard deviations.
The first and second uncertainties are statistical and systematic, respectively.
Using the world-average value of the branching fraction for $\Lambda_c^+ \to \Lambda \pi^+ \pi^0$, we obtain the branching fractions $\mathcal{B}(\Lambda_c^+ \to \Lambda K^+ \pi^0)=(1.29\pm 0.05\pm 0.05\pm 0.06)\times10^{-3}$ and $\mathcal{B}(\Lambda_c^+ \to \Lambda K^+ \eta)=(6.20\pm 0.25\pm 0.31\pm 0.30)\times10^{-4}$, where the third uncertainty arises from the uncertainty in the world-average value of $\mathcal{B}(\Lambda_c^+ \to \Lambda \pi^+ \pi^0)$.
The measured $\mathcal{B}(\Lambda_c^+\to\Lambda K^+\eta)$ is significantly larger than predictions based on $SU(3)$ flavor symmetry.
		
	\end{abstract}

	\maketitle

	\section{Introduction}

The hadronic decays of charmed baryons provide an important platform for investigating the intricate interplay between weak and strong interactions within the standard model.
Although the factorization approach has been successfully applied to charmed meson decays, nonfactorizable contributions play an essential role in charmed baryon decays due to the absence of helicity and color suppression~\cite{theory:review:Cheng:2021qpd}.
Over the past few decades, significant efforts have been devoted to developing theoretical approaches that overcome the limitations of factorization. Among them, the $SU(3)_F$ flavor symmetry approach has emerged as a useful framework~\cite{Bjorken:1988ya,Cheng:1991sn,Cheng:1993gf,Zenczykowski:1993hw,Cheng:1995fe,Fayyazuddin:1996iy,Dhir:2015tja,Cheng:2018hwl}.
Its applicability has been established in hadronic two-body decays and semileptonic decays of charmed baryons~\cite{Savage:1989qr,Savage:1991wu,Pirtskhalava:2011va,Grossman:2012ry,Lu:2016ogy,Geng:2017esc,Geng:2017mxn,Geng:2018plk,Geng:2018bow,Hsiao:2019yur,Geng:2019bfz,Jia:2019zxi}.
The hadronic three-body charmed baryon decays remain theoretically challenging due to the complex decay dynamics and possible intermediate resonant contributions. 
Among them, singly Cabibbo-suppressed (SCS) modes are particularly poorly explored, 
making precise measurements and searches for SCS decays essential for understanding the underlying decay mechanism.

The SCS decay $\Lambda_c^+ \to \Lambda K^+ \pi^0$ was first observed by the BESIII Collaboration in 2024~\cite{Exp:Lc2lbdkpi:BESIII:2023sdr}.
(Charge-conjugate processes are implied throughout this paper.)
Its corresponding branching fraction was measured to be $\mathcal{B}(\Lambda_c^+ \to \Lambda K^+ \pi^0 )=(1.49\pm0.27\pm0.05\pm0.08)\times 10^{-3}$, using the Cabibbo-favored decay $\Lambda_c^+\to\Lambda\pi^+\pi^0$ as a reference mode. 
The first and second uncertainties denote the statistical and systematic uncertainties, respectively, while the third uncertainty arises from the uncertainty in the world-average value of $\mathcal{B}(\Lambda_c^+\to\Lambda\pi^+\pi^0)$.
The result is significantly lower than the theoretical predictions of $(4.5\pm0.8)\times 10^{-3}$ in Ref.~\cite{Theory:Predict1:Geng:2018upx} and $(3.5\pm0.6)\times 10^{-3}$ in Ref.~\cite{Thoery:Predict2:Cen:2019ims}.
In these two early calculations, only the color-antisymmetric part of the effective Hamiltonian was considered.
These calculations assumed a non-resonant, $S$-wave-dominant pseudoscalar meson-pair configuration (i.e., with relative orbital angular momentum $L=0$),  and performed a global fit with 12 parameters using the available measured branching fractions of hadronic three-body decays of charmed baryons at that time~\cite{Theory:Predict1:Geng:2018upx,Thoery:Predict2:Cen:2019ims}.
Later, Reference~\cite{Theory:Predict3:Geng:2024sgq} extended the analysis by including the complete effective Hamiltonian, i.e., both the color-antisymmetric and color-symmetric parts.
The extended fit introduces four additional free parameters.
The resulting fit, with the BESIII measurement of $\mathcal{B}(\Lambda_c^+ \to \Lambda K^+ \pi^0)$~\cite{Exp:Lc2lbdkpi:BESIII:2023sdr} as one of the inputs, yields $\mathcal{B}(\Lambda_c^+ \to \Lambda K^+ \pi^0)=(1.36 \pm 0.25)\times 10^{-3}$~\cite{Theory:Predict3:Geng:2024sgq}.

The decay $\Lambda_c^+ \to \Lambda K^+ \eta$ has not yet been observed. 
Within the $SU(3)_F$ framework, it proceeds through the same topological diagrams as $\Lambda_c^+ \to \Lambda K^+ \pi^0$, but with a different isospin final state and much smaller available phase space.
The branching fraction, $\mathcal{B}(\Lambda_c^+ \to \Lambda K^+ \eta)$, has been predicted to be $(1.9\pm0.6)\times10^{-4}$~\cite{Theory:Predict1:Geng:2018upx}, $(0.45\pm0.10)\times10^{-4}$~\cite{Thoery:Predict2:Cen:2019ims}, and more recently $(0.59\pm0.24)\times10^{-4}$~\cite{Theory:Predict3:Geng:2024sgq}, reflecting differences among the theoretical approaches.
The sizable spread among these predictions highlights the need for experimental measurements to discriminate between the theoretical approaches.

In this paper, we report measurements of the branching fractions of SCS decays, $\Lambda_c^+ \to \Lambda K^+ \pi^0$ and $\Lambda_c^+ \to \Lambda K^+ \eta$, relative to the Cabibbo-favored decay $\Lambda_c^+ \to \Lambda \pi^+ \pi^0$, using data corresponding to integrated luminosities of 980\invfb and 570\invfb, collected with the Belle and Belle~II detectors, respectively.

\section{Data Samples}

The Belle experiment operated at the KEKB~\cite{kekb,kekb1} asymmetric-energy $e^{+}e^{-}$ collider with electron- and positron-beam energies of 8.0\gev and 3.5\gev, respectively, while the Belle~II experiment operates at its successor, SuperKEKB~\cite{skekb}, with corresponding beam energies of 7.0\gev and 4.0\gev.
The Belle data were collected at or near the $\Upsilon(nS)$ $(n=1,2,3,4,5)$ resonances, whereas the Belle II data were collected  at or near the $\Upsilon(4S)$ resonance and at a center-of-mass energy of $\sqrt{s} = 10.75$\gev.

The Belle detector had a cylindrical symmetry around the beamline, with the $z$-axis defined as the direction of the electron beam. The Belle detector comprised six sub-detectors, listed in the following from innermost to outermost: the silicon vertex detector (SVD), the central drift chamber CDC, the aerogel Cherenkov counter (ACC), the time-of-flight (TOF) scintillation counter, the electromagnetic calorimeter (ECL), and the \KL and muon (KLM) detector. The ECL consisted of a cylindrical barrel region surrounding the beamline and forward and backward endcap regions at either end, with the forward endcap located in the positive $z$ direction. A solenoid produced a 1.5~T magnetic field through the five innermost detectors. Further details are provided in Ref.~\cite{Belle:detector}.

The Belle~II detector is an upgraded version of Belle detector, featuring  several new subsystems and substantial upgrades to others.
Charged-particle trajectories (tracks) are reconstructed using a two-layer silicon pixel detector, surrounded by a four-layer double-sided silicon-strip detector and a new large-radius, small-cell CDC. 
Of the full data sample, 430~\(\mathrm{fb}^{-1}\) were collected while only 15\% of the second pixel layer had been installed.
Outside the CDC, a time-of-propagation detector and an aerogel ring-imaging Cherenkov detector cover the barrel and forward endcap regions, respectively. 
The CsI(Tl) crystal ECL, solenoid, and the iron flux-return yoke from the Belle detector are reused in the Belle~II detector. The ECL readout electronics have been upgraded, and the instrumentation in the flux-return yoke to identify $K_L^0$ mesons and muons has been replaced.

The $z$ axis of the laboratory frame is defined as the detector solenoid axis, with the positive direction along the electron beam direction.
The polar angle $\theta$ and the transverse plane are defined relative to this axis.

Simulated events are used to optimize the event selection criteria and develop the analysis strategy, including background suppression and signal extraction. 
Continuum $e^+e^-\to \qqbar$ ($\quark= \uquark, \dquark, \squark, \cquark$) events are generated by \kkmc~\cite{kkmc} software, interfaced with {\mbox{\textsc{Pythia}}\xspace}~\cite{pythia1,pythia2} for the subsequent hadronization.
The decays of unstable particles are simulated with \evtgen~\cite{evtgen}, using the known branching fractions~\cite{pdg}.
The final-state radiation is simulated by the \photos~\cite{photos} package.
In simulations, $\Lambda_c^+$ baryons are produced in $e^+ e^- \to c \bar c$ processes. The decays of the two signal modes, $\Lambda_c^+ \to \Lambda K^+\pi^0 $ and $\Lambda_c^+\to\Lambda K^+\eta$, and the reference mode $\Lambda_c^+ \to \Lambda \pi^+\pi^0 $ are simulated  
according to a phase-space model.
Exclusive simulated samples of $\Lambda_c^+\to\Sigma^0\pi^+\pi^0$, $\Lambda_c^+\to\Sigma^0 K^+\pi^0$, $\Lambda_c^+\to\Sigma^0 K^+\eta$, and $\Lambda_c^+\to\Lambda \pi^+ \eta$ are generated to study the corresponding background contributions. 
For each decay channel,  two million events are generated for each of the Belle and Belle~II simulations.
The detector response is simulated using the {\mbox{\textsc{Geant3}}\xspace}~\cite{geant3} software package for Belle and {\mbox{\textsc{Geant4}}\xspace}~\cite{geant4} for Belle~II.
We use the Belle~II analysis software framework (basf2)~\cite{basf2, basf2-zenodo} to reconstruct and analyze both Belle and Belle~II data.
The Belle data are converted to the Belle~II format for basf2 compatibility using the B2BII framework~\cite{b2bii}.

\section{Event Selection Criteria}

We reconstruct the signal decays  $\Lambda_c^+\to\Lambda K^+ \pi^0$ and $\Lambda_c^+\to\Lambda K^+ \eta$ and the reference decay $\Lambda_c^+ \to \Lambda \pi^+ \pi^0$, where $\Lambda\to p\pi^-$, $\pi^0\to\gamma\gamma$, and $\eta\to\gamma\gamma$.
The  intermediate particles $\Lambda$, $\pi^+$, $K^+$, $\pi^0$, $\eta$ are pre-selected from offline data as follows.

Charged-particle identification (PID) is based on likelihoods constructed from the responses of the detector subsystems. 
At Belle, the likelihood ${\cal L}(i)$ for each particle hypothesis $i$ is formed using information from the CDC, ACC, and TOF~\cite{b1PID:Nakano:2002jw}. 
The $\pi$, $K$, and $p$ candidates are identified using a binary discriminator ${\cal P}(h\colon h^\prime)={\cal L}(h)/[{\cal L}(h)+{\cal L}(h^{\prime})]$, where $h, h^{\prime}$ denote a $\pi$, $K$, or $p$.
At Belle~II, the PID likelihoods are constructed using information from the CDC, TOP, ARICH, ECL, and KLM~\cite{b2hid:Belle-II:2025tpe}.
For $\pi$ and $K$ candidates, the binary discriminators ${\cal P}(\pi\colon K)$ and ${\cal P}(K\colon \pi)$ are used, while
the discriminator ${\cal P}(h)={\cal L}(h)/[\sum_i{\cal L}(i)]$ is  used for the proton candidates, where $i$ denotes a $\pi$, $K$, $p$, $e$, $\mu$, or deuteron.

The $\Lambda$ candidates are reconstructed from a proton and an oppositely charged pion candidate. 
For the Belle data, the proton candidates are required to satisfy $\mathcal{P}(p:\pi)>0.1$.
Momentum-dependent requirements on variables describing the displaced decay topology, as detailed in Ref.~\cite{Belle:2020xku}, are applied. 
For the Belle~II data, the discriminator $\mathcal{P}(p)> 0.1$ is used. 
We apply a boosted decision tree (BDT) classifier trained using kinematic properties of the $\Lambda$ candidate and its decay products, together with the candidates flight lengths.
For both data samples, the invariant mass of the $\Lambda$ candidate is required to be within $\pm3$\mevcc of the known mass~\cite{pdg}, corresponding to approximately 2.5 times the mass resolution ($\sigma$). 

Charged particles not used in the $\Lambda$ reconstruction are required to have transverse and longitudinal distances of closest approach to the interaction point of less than 2\cm and 4\cm, respectively. 
We retain pion and kaon candidates with requirements ${\cal P}(\pi \colon K) > 0.6$ and ${\cal P}(K\colon\pi) > 0.6$, respectively, for both the Belle and Belle~II data samples. The resulting PID efficiencies range from 91\% to 94\%, depending on particle species and data samples,
while the corresponding misidentification rates range from 4\% to 9\%, determined using the control sample $D^{*+}\to D^0(\to K^-\pi^+)\pi^+$ in bins of momentum, polar angle, and charge~\cite{b1PID:Nakano:2002jw,b2hid:Belle-II:2025tpe}.

Photon candidates are identified as ECL energy deposits within the CDC acceptance that are not associated with any charged particle. 
The neutral meson candidates $\pi^0$ and $\eta$ are reconstructed from pairs of photons within the invariant mass windows of 120\mevcc $< M(\gamma\gamma) <145$\mevcc and 520\mevcc $< M(\gamma\gamma) < 570$\mevcc, respectively, corresponding to ($-3\sigma$, $2\sigma$) and $\pm2\sigma$ around  the known masses~\cite{pdg}. The window sizes are chosen according to the signal shapes and background levels.

The selected $\Lambda$, $\pi^+$, $K^+$, $\pi^0$, and $\eta$ candidates are combined to reconstruct $\Lambda_c^+$ candidates.
The subsequent selection requirements for the $\Lambda_c^+$ candidates are optimized 
by maximizing the figure-of-merit, $N_{\rm sig}/\sqrt{N_{\rm sig}+N_{\rm bkg}}$, for the $\Lambda_c^+\to\Lambda K^+\pi^0$ mode, and the Punzi figure-of-merit~\cite{punzi}, $\varepsilon/(a/2+\sqrt{N_{\rm bkg}})$, for the $\Lambda_c^+\to\Lambda K^+ \eta$ mode, since the branching fraction ${\cal B}(\Lambda_c^+\to\Lambda K^+\eta)$ has not yet been measured and its theoretical predictions span a wide range~\cite{Theory:Predict1:Geng:2018upx,Thoery:Predict2:Cen:2019ims,Theory:Predict3:Geng:2024sgq}.
The parameter $a$ is set to 5, corresponding to the target $5\sigma$ sensitivity for the $\Lambda_c^+\to\Lambda K^+\eta$ search.
Here, $N_{\rm sig}$ is the expected signal yield, estimated from the reconstruction efficiency, the measured branching fraction $\mathcal{B}(\Lambda_c^+\to\Lambda K^+\pi^0)$~\cite{Exp:Lc2lbdkpi:BESIII:2023sdr}, and the measured yield of the reference mode in data together with its known branching fraction~\cite{pdg}.
The reconstruction efficiency, $\varepsilon$, and the expected background yield in the signal region, $N_{\rm bkg}$, are determined, respectively, from the signal simulation and continuum $e^+ e^- \to q \bar q$ simulated events, hereafter referred to as inclusive simulated samples. The signal region used to optimize the selections is defined by the reconstructed $\Lambda_c^+$ invariant mass as $[2.26,\,2.31]$\gevcc for the $\Lambda^+_c \to \Lambda \pi^+(K^+) \pi^0$ channels and $[2.27,\,2.30]$\gevcc for the $\Lambda^+_c \to \Lambda K^+  \eta$ channel, corresponding to approximately $\pm3\sigma$.
The sideband regions are defined as $[2.19,\,2.24]$ and $[2.32,\,2.37]$\gevcc for the $\Lambda^+_c \to \Lambda \pi^+(K^+) \pi^0$ channels, and $[2.23,\,2.26]$ and $[2.31,\,2.34]$\gevcc for the $\Lambda^+_c \to \Lambda K^+ \eta$ channel.

A global decay-chain vertex fit is applied to each selected $\Lambda_c^+$ candidate using the TreeFit algorithm~\cite{b2fit}, where the masses of the $\Lambda$, $\pi^0$, and $\eta$  are constrained to their known values~\cite{pdg}.
The $\chi^2$ per degree of freedom of the fit is required to be less than 10.
We require the scaled momentum of the $\Lambda_c^+$ candidate, $x_p \equiv p^*c/\sqrt{s/4-M^2(\Lambda^+_c)c^4}$, to exceed 0.55 to suppress backgrounds from $B$ meson decays and combinatorial backgrounds. Here, $p^*$ and $M(\Lambda^+_c)$ are the momentum in the center-of-mass frame and the known mass~\cite{pdg} of $\Lambda_c^+$ , respectively, and $s$ is the square of the $e^+e^-$ center-of-mass energy.

To suppress contributions from misreconstructed  photons and beam background, dedicated BDT classifiers are applied to photon candidates in the Belle and Belle~II analyses, as described in Ref.~\cite{photon:mva}.
The energies of photons are required to exceed 70\mev and 200\mev, respectively, to form the $\pi^0$ and $\eta$ candidates.
The momenta of the $\Lambda$, $\pi^0$, and $\eta$ candidates in the center-of-mass frame are required to exceed 1\gevcc, 0.36\gevcc, and 0.5\gevcc, respectively, to suppress combinatorial backgrounds.
To suppress background from non-signal $\pi^0$  mesons, events are rejected if the invariant mass of the signal photon combined with any other photon in the event is within 5\mevcc and 10\mevcc of the known $\pi^0$ mass~\cite{pdg},  for the $\Lambda_c^+\to\Lambda\pi^+(K^+)\pi^0$ and $\Lambda_c^+\to\Lambda K^+\eta$ modes, respectively. This veto targets cases in which one daughter photon from a non-signal $\pi^0$ is misreconstructed as the signal photon.
The $\pi^0$ veto retains approximately 89\%~(66\%) of the signal candidates in the $\pi^0$ selection and 90\%~(60\%) in the $\eta$ selection for Belle (Belle~II). The corresponding background rejection rates are 23\%~(47\%) and 38\%~(70\%), respectively.
The lower signal retention in Belle~II is attributed to the higher photon background level, which increases the probability of forming a $\pi^0$ candidate.

To suppress contributions from the two-body decay $\Lambda_c^+\to\Xi^0 K^+$ followed by $\Xi^0\to\Lambda\pi^0$,
we reject candidates with a  $\Lambda\pi^0$ invariant mass within approximately $\pm3\sigma$ of the known $\Xi^0$ mass~\cite{pdg}.  This requirement retains more than 98\% of the signal three-body decays.
Background contributions from inclusive $\Sigma^0$, $\Xi^-$, and $\Xi^0$ hyperons decaying to $\Lambda$ are identified using the TopoAna algorithm~\cite{topo}.
To suppress these backgrounds, we combine the selected $\Lambda$ candidate with particles reconstructed from the rest of the event to form $\Sigma^0$, $\Xi^-$, and $\Xi^0$ candidates, and reject events in which the invariant mass of the $\Lambda\gamma$, $\Lambda\pi^-$, or $\Lambda\pi^0$ combination lies within approximately $\pm3\sigma$ of the corresponding known hyperon mass~\cite{pdg}.
The photon candidates used in the $\Sigma^0$ and $\Xi^0$ vetoes are required to satisfy the photon BDT selection and to have energies exceeding 70\mev and 50\mev, respectively. The $\pi^-$ candidate used in the $\Xi^-$ veto is required to satisfy the same pion identification requirement applied to the signal pions
and the $\Xi^-\to\Lambda\pi^-$ candidate is required to have a flight-distance significance greater than 2.
These vetoes reduce the reconstruction efficiency of signal modes by 13\%--17\%, 0.2\%--0.5\%, and 5\%--7\% for the $\Sigma^0$, $\Xi^-$, and $\Xi^0$ vetoes, respectively, depending on the decay channel and data samples,  while rejecting 45\%--55\%, 44\%--50\%, and 20\%--33\% of the corresponding background events.

After all reconstruction and selection requirements, the average $\Lambda_c^+$ candidate multiplicity is 1.03 per event for both Belle and Belle~II data and simulation. 
All $\Lambda_c^+$ candidates are retained in events with multiple candidates.

\section{Yield Extraction in Data}

\begin{figure*}[t]	
	\centering
	\includegraphics[width=7.0cm]{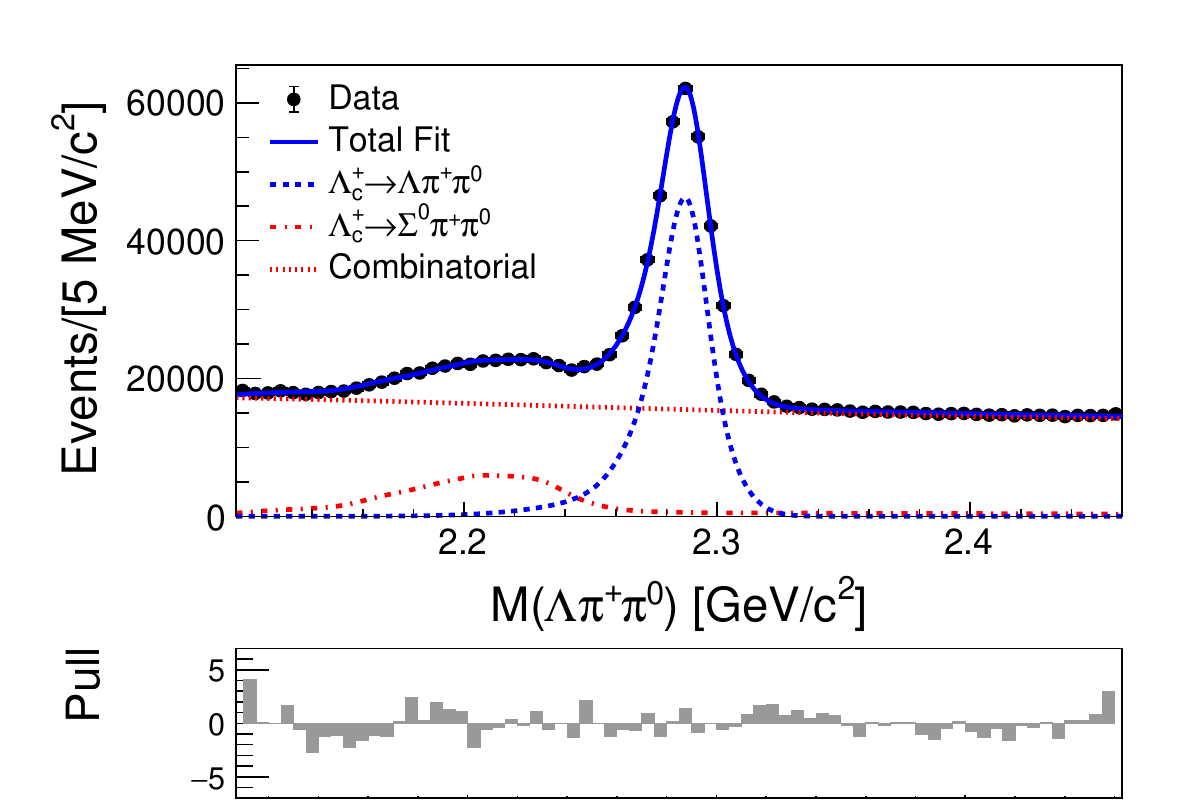}		 \put(-200,130){\bf (a)} \put(-100,130){\bf Belle}
	\put(-70,150){{\bf Belle + Belle II} $\rm \bf \int Ldt=1.55$\invab}
	\includegraphics[width=7.0cm]{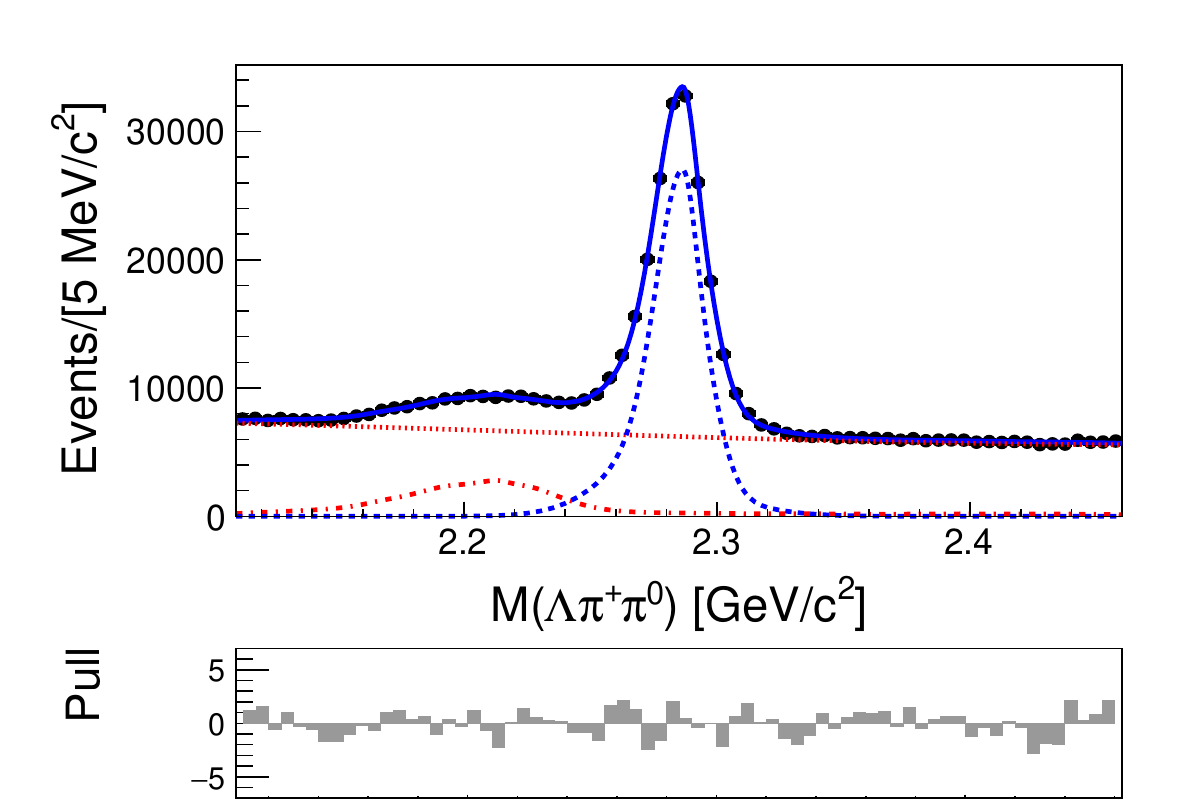}	
	\put(-100,130){\bf Belle II}

	\includegraphics[width=7.0cm]{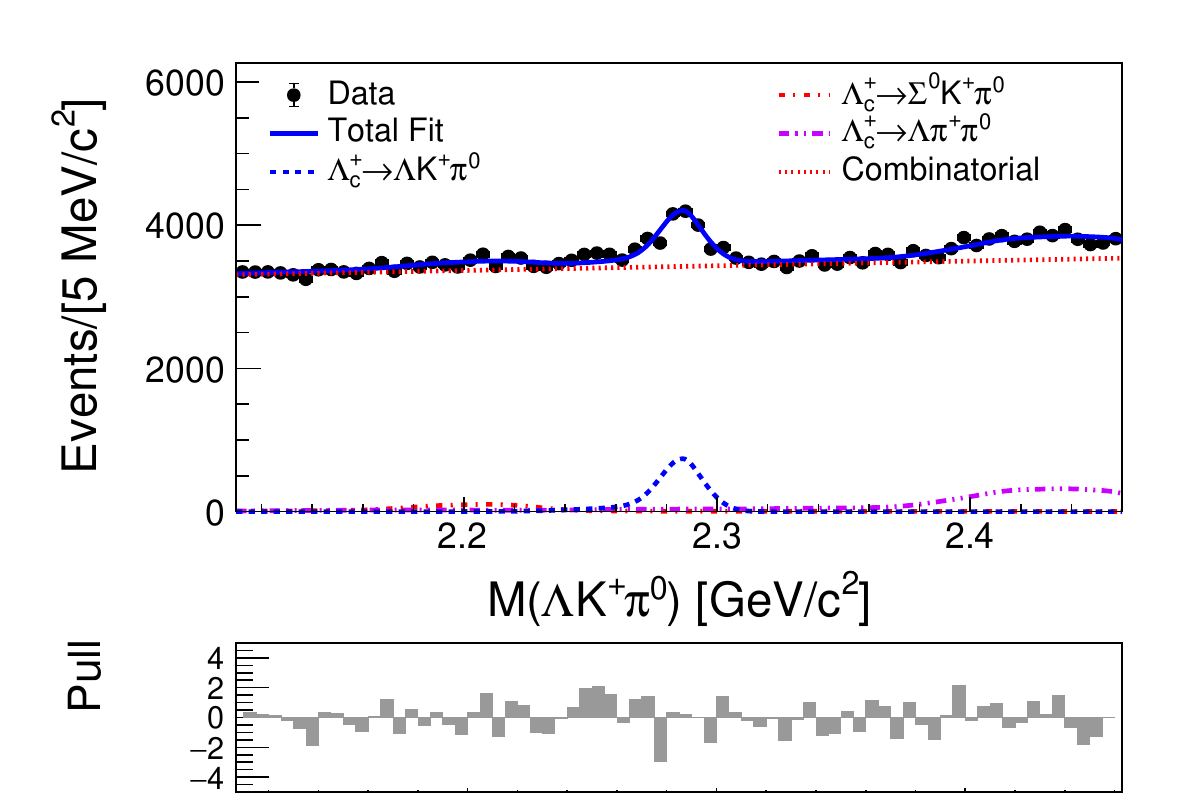}		 \put(-200,130){\bf (b)} 
	\includegraphics[width=7.0cm]{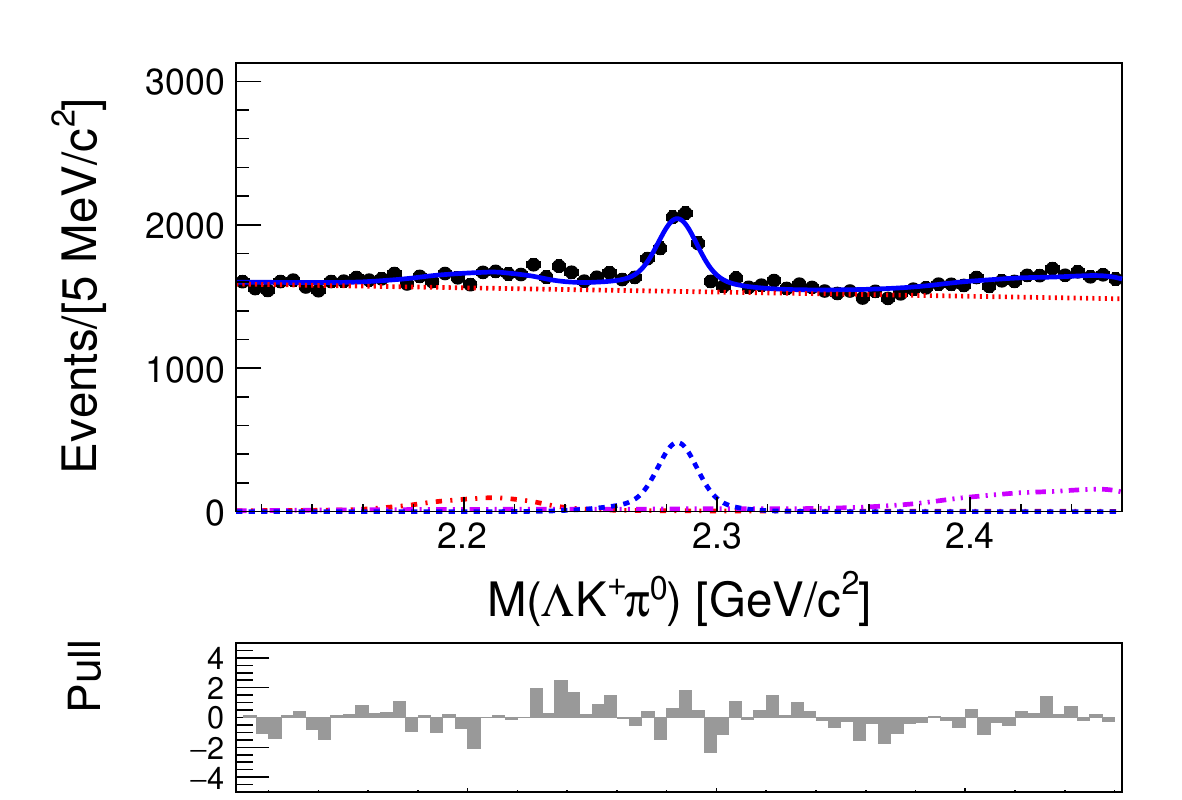}

	\includegraphics[width=7.0cm]{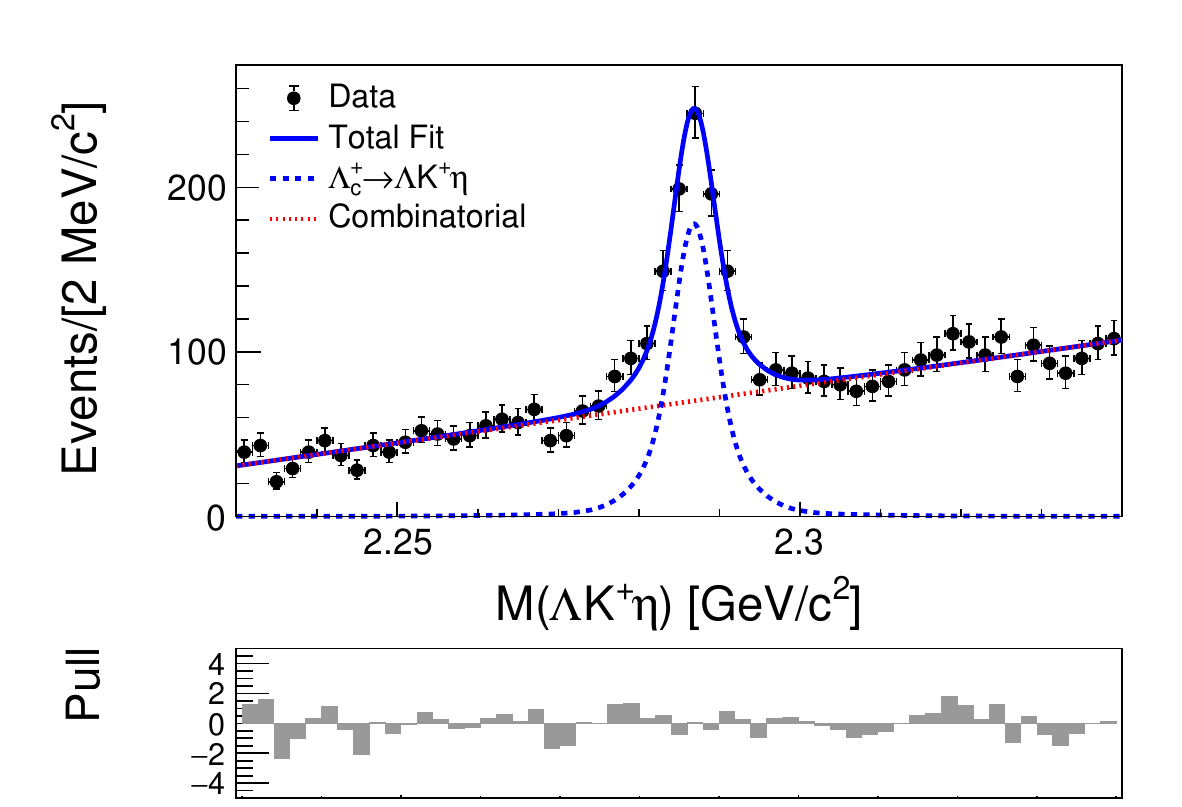}		 \put(-200,130){\bf (c)} 
	\includegraphics[width=7.0cm]{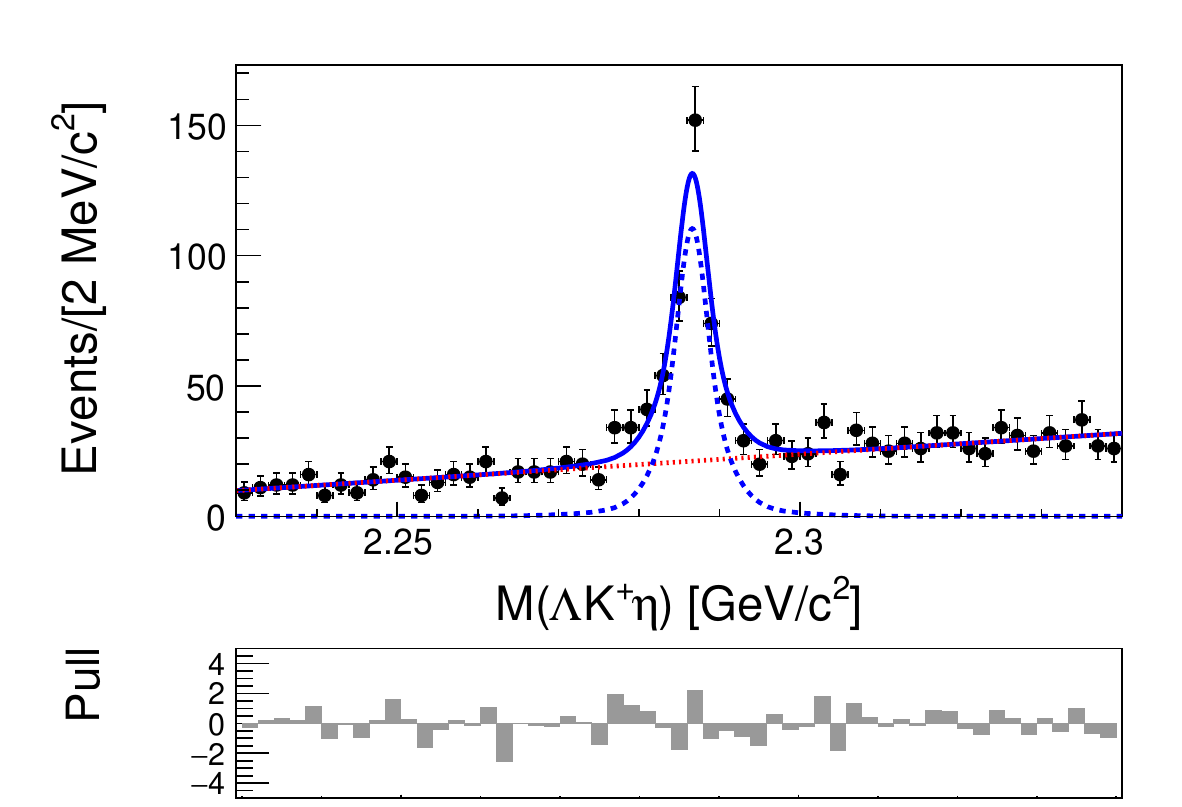}	

	\caption{Reconstructed invariant-mass distributions of $\Lambda_c^+$ candidates in (a) $\Lambda\pi^+\pi^0$, (b) $\Lambda K^+\pi^0$, and (c) $\Lambda K^+\eta$ channels for the Belle (left) and Belle~II (right) data samples. Data are shown as black points with error bars, while the solid blue curves show the total fit to data. The dashed blue, dash-dotted red, dash-double-dotted violet, and dotted red curves represent the signal, feed-down background, misidentification background, and combinatorial background, respectively.
    The gray bars show the pull distributions of the fits, defined as $(N_{\rm data}-N_{\rm fit})/\sigma_{\rm data}$. Here, $N_{\rm data}$, $N_{\rm fit}$, and $\sigma_{\rm data}$ denote the number of data events, the fitted yield, and the uncertainty on $N_{\rm data}$, respectively, in each displayed bin.}
	\label{fig:Mlambdac:data}
\end{figure*}

The invariant mass distributions of the reconstructed $\Lambda_c^+\to\Lambda \pi^+\pi^0$, $\Lambda_c^+\to\Lambda K^+\pi^0$, and $\Lambda_c^+\to\Lambda K^+ \eta$ decays are shown in Fig.~\ref{fig:Mlambdac:data}. 
Two peaking backgrounds are observed near the $\Lambda_c^+$ peak in the simulated samples.
One background component forms a broad structure on the low-mass side of the $\Lambda_c^+$ peak due to the missing photon from the $\Sigma^0$ decay. It originates from the $\Lambda_c^+\to\Sigma^0\pi^+\pi^0$ and $\Lambda_c^+\to\Sigma^0 K^+\pi^0$ decays in the $\Lambda_c^+\to\Lambda\pi^+\pi^0$ and $\Lambda_c^+\to\Lambda K^+\pi^0$ channels, respectively.
Another broad structure on the high-mass side of the $\Lambda_c^+\to \Lambda K^+\pi^0$ signal originates from the Cabibbo-favored reference channel $\Lambda_c^+\to\Lambda\pi^+\pi^0$,  where the pion is misidentified as a kaon. 
For the $\Lambda_c^+\to\Lambda K^+\eta$ channel, both background components are located more than $10\sigma$ away from the signal peak and therefore have negligible impact on the signal extraction.
The fit ranges are hence set at approximately $\pm20\sigma$ for $\Lambda_c^+\to\Lambda\pi^+\pi^0$ and $\Lambda_c^+\to\Lambda K^+\pi^0$ channels to include those broad structures, and approximately $\pm10\sigma$ for the $\Lambda_c^+\to\Lambda K^+\eta$ mode.

\begin{figure*}[]	
	\centering    
	\includegraphics[width=5.0cm]{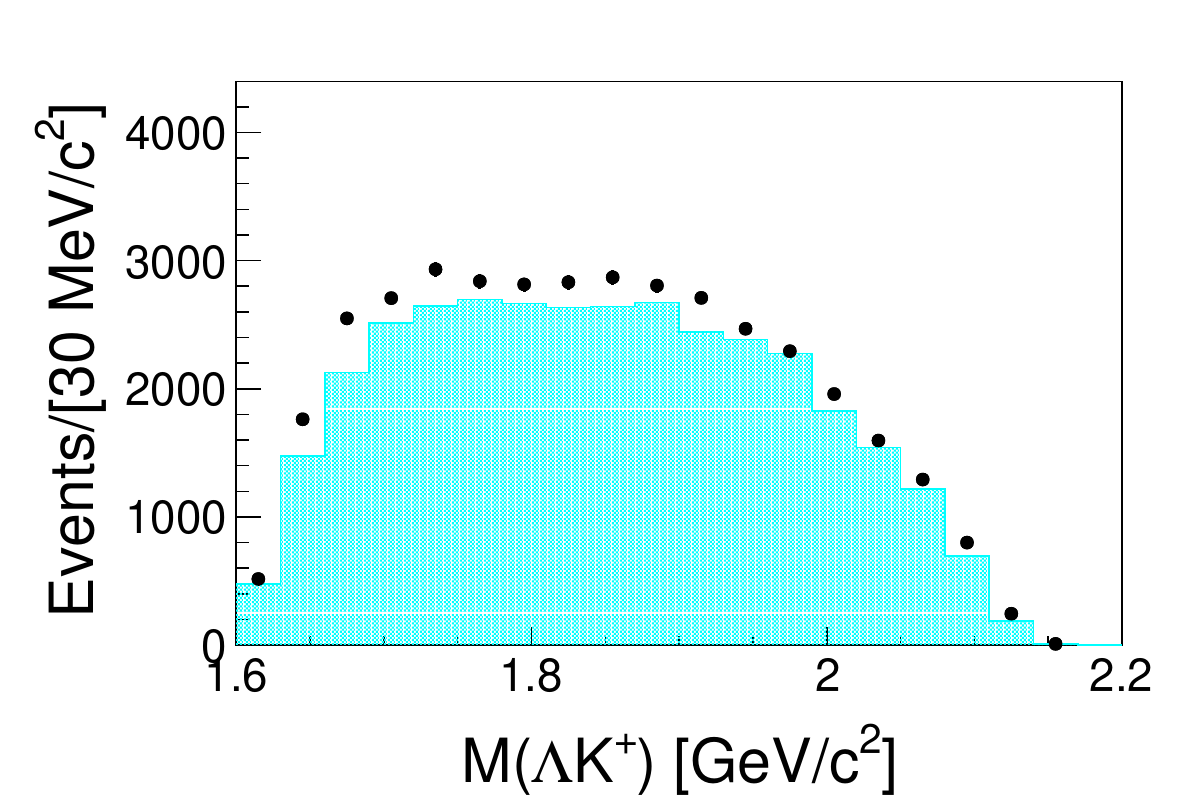} \put(-180,70){\bf Belle} \put(-150,90){\bf (a)} 
	\put(-10,100){{\bf Belle + Belle II} $\rm \bf \int Ldt=1.55$\invab}
	\includegraphics[width=5.0cm]{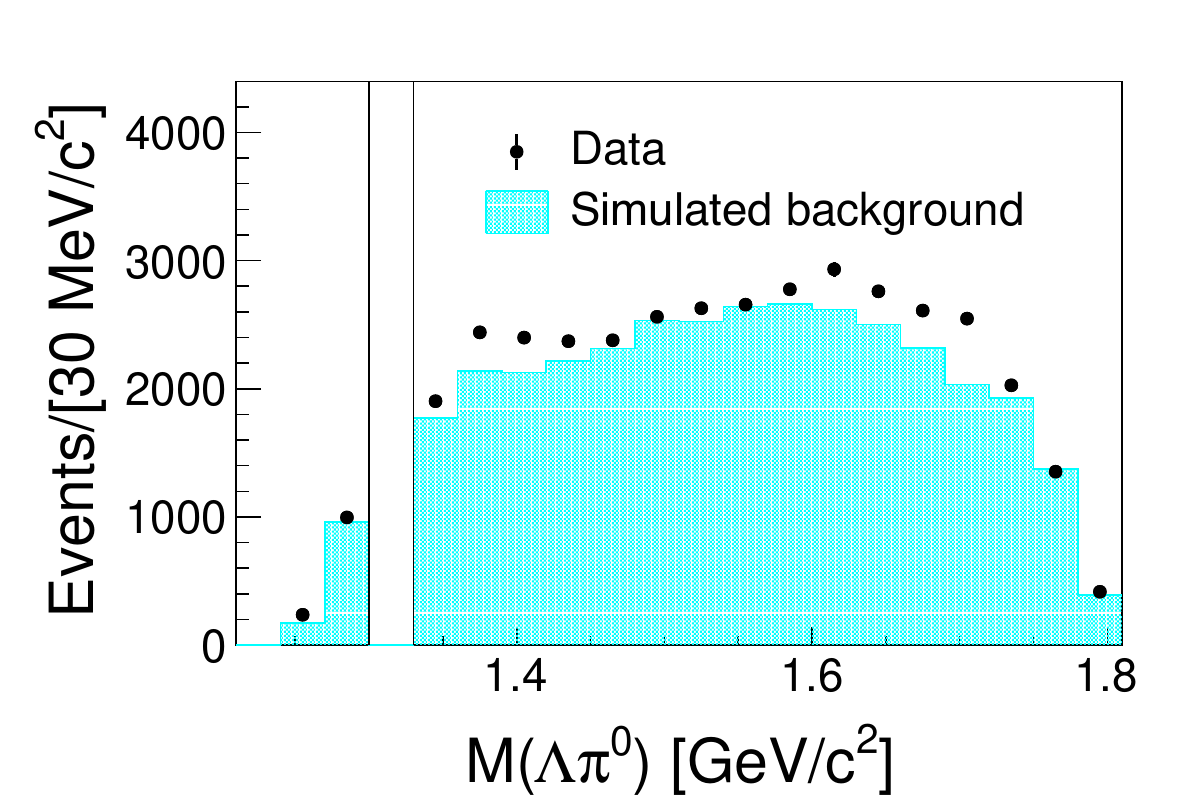} 
	\includegraphics[width=5.0cm]{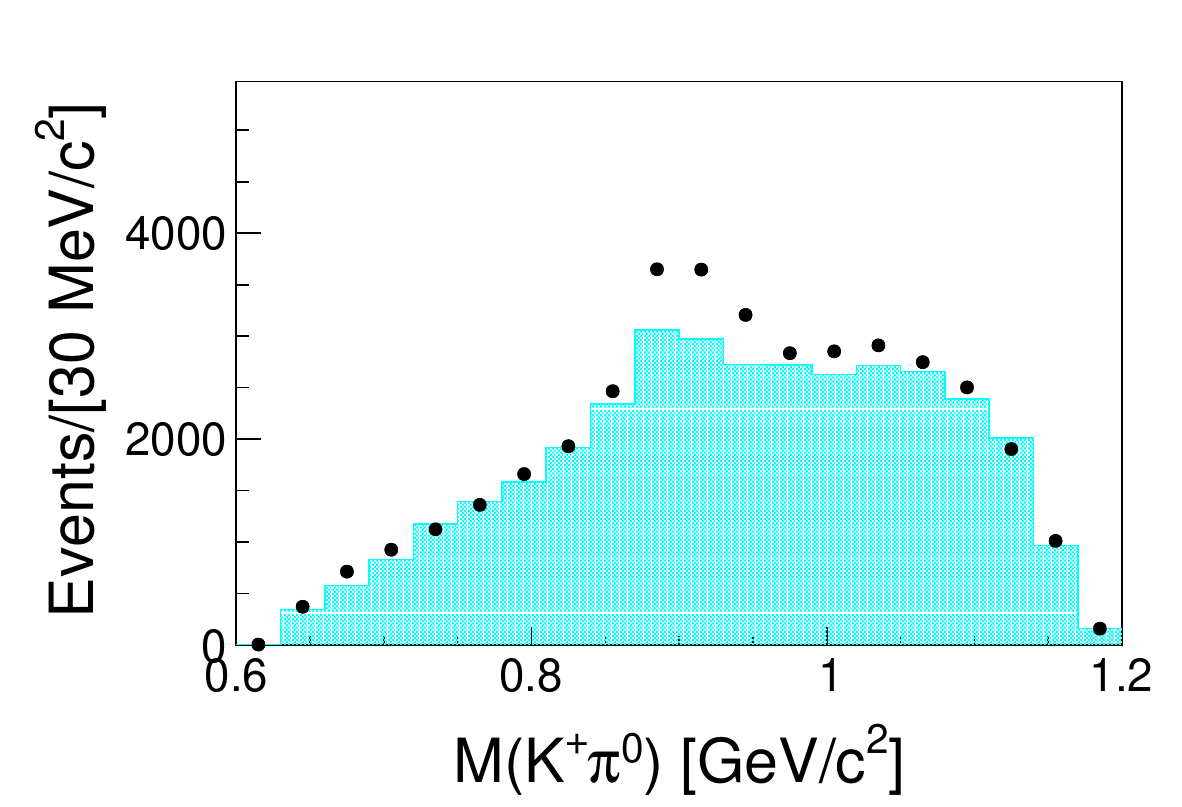} 	
	
	\includegraphics[width=5.0cm]{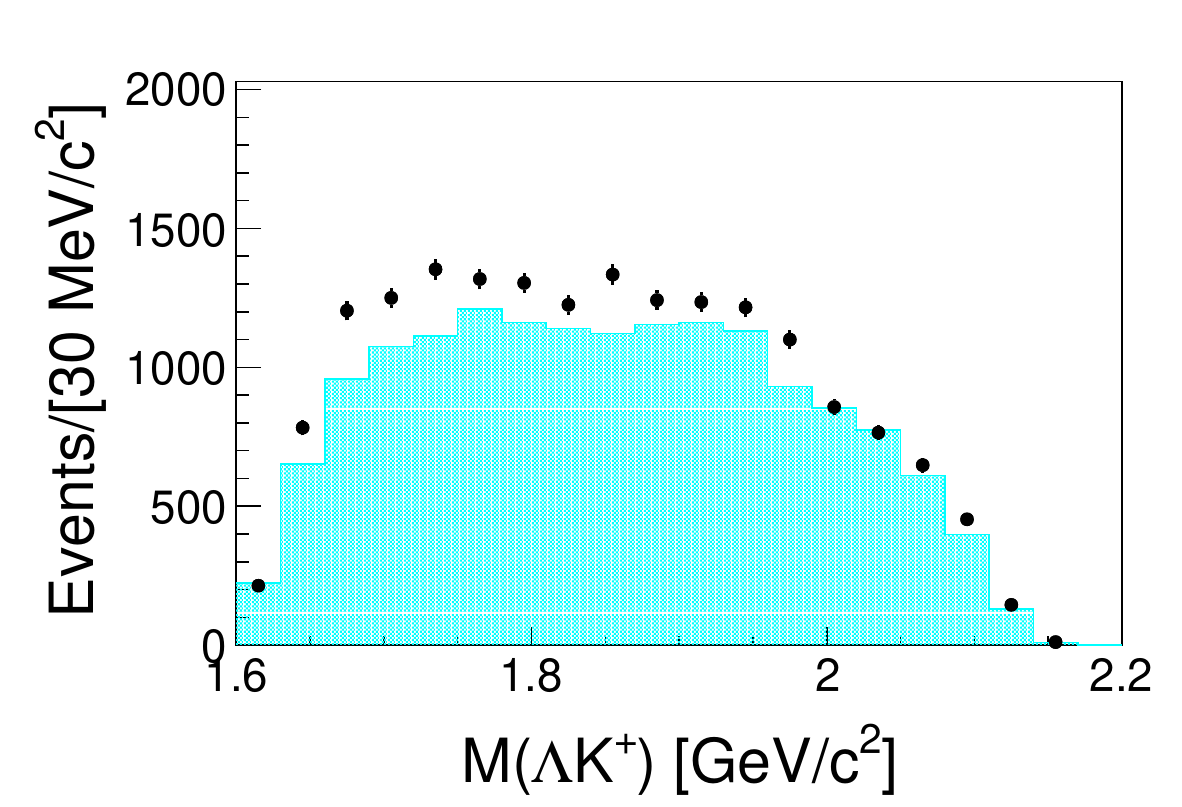}\put(-180,70){\bf Belle II}
	\includegraphics[width=5.0cm]{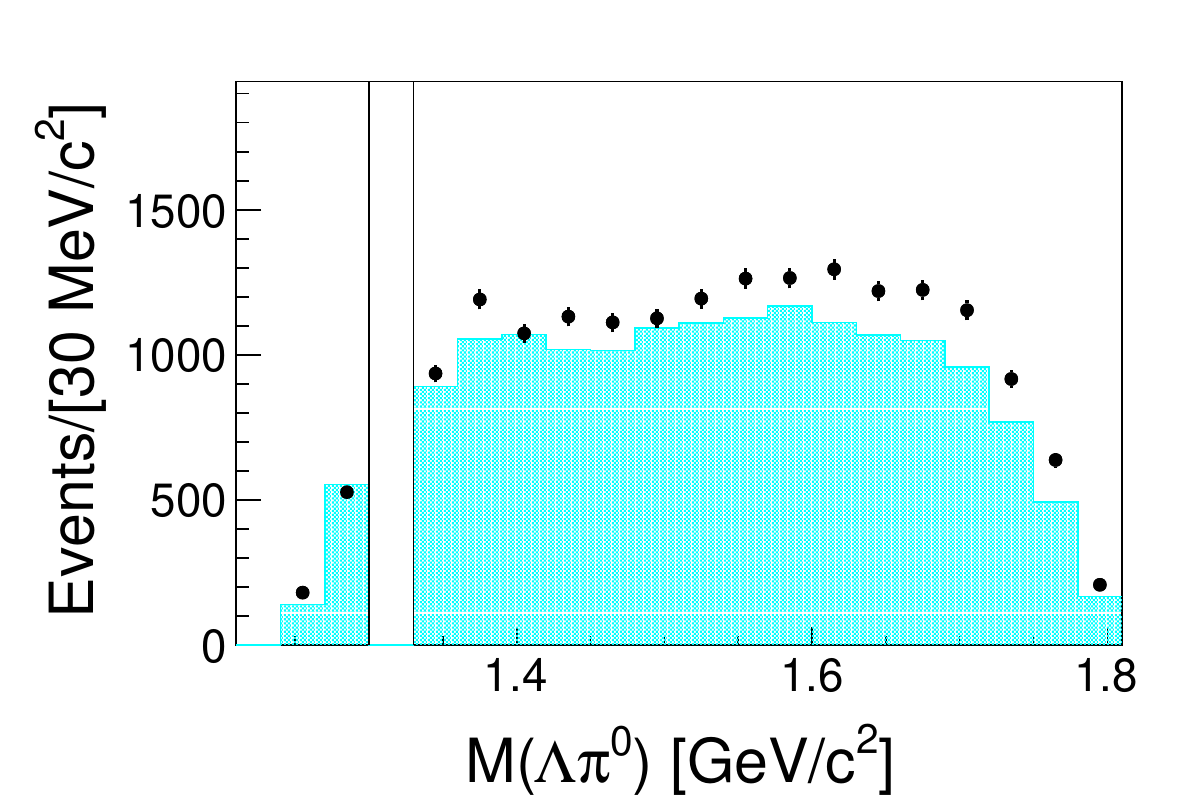}
	\includegraphics[width=5.0cm]{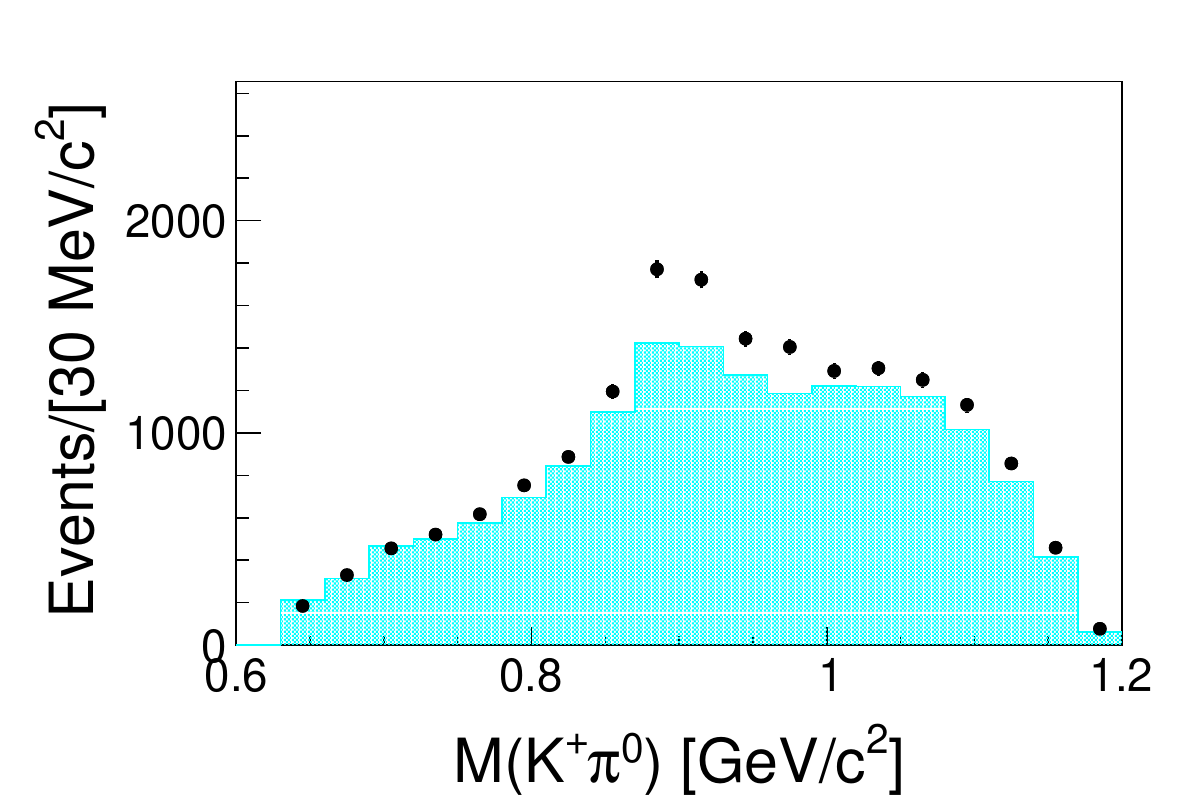}
	
	\includegraphics[width=5.0cm]{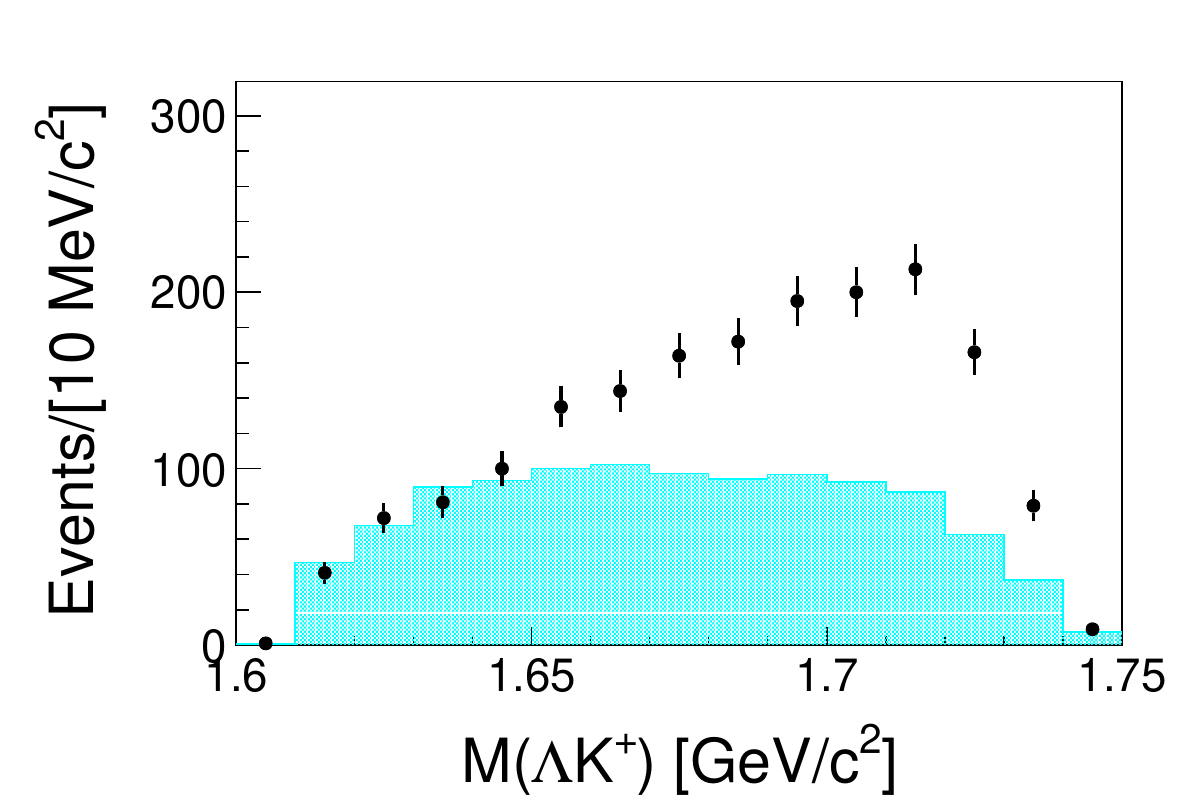} \put(-180,70){\bf Belle} \put(-150,90){\bf (b)} 
    \includegraphics[width=5.0cm]{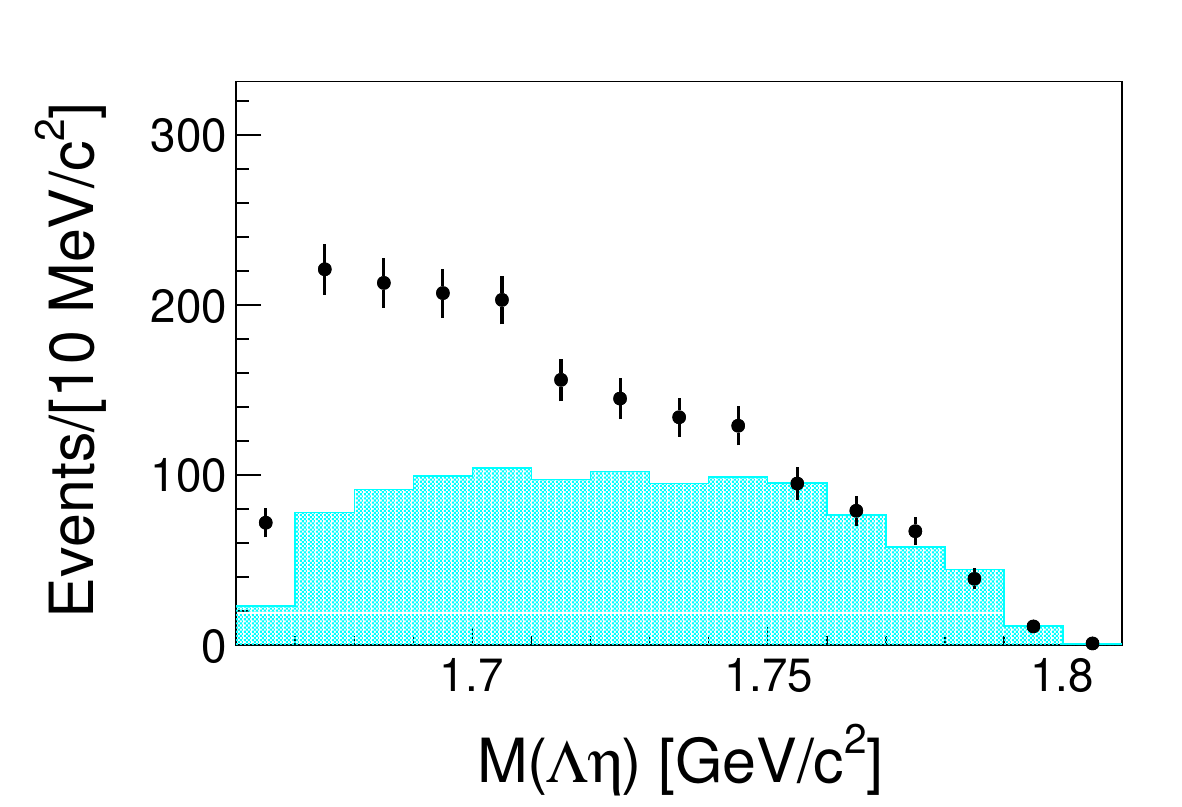} 
	\includegraphics[width=5.0cm]{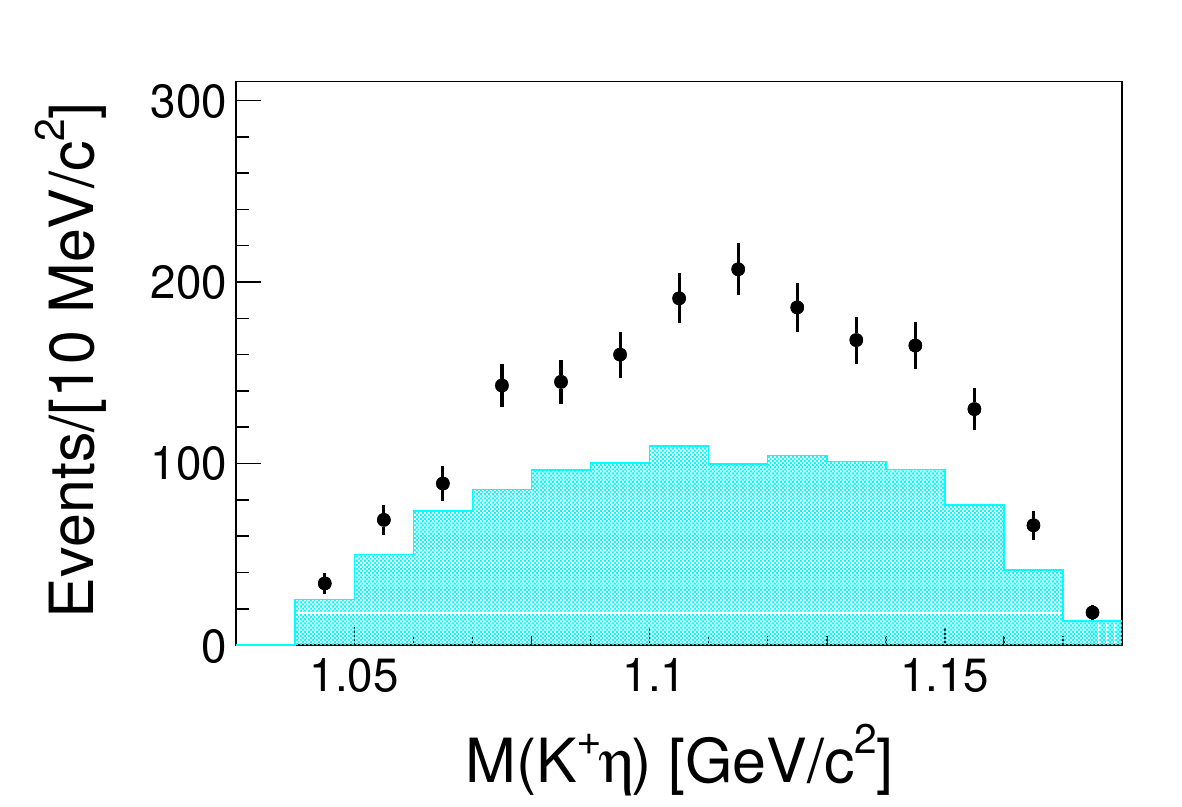} 	
	
	\includegraphics[width=5.0cm]{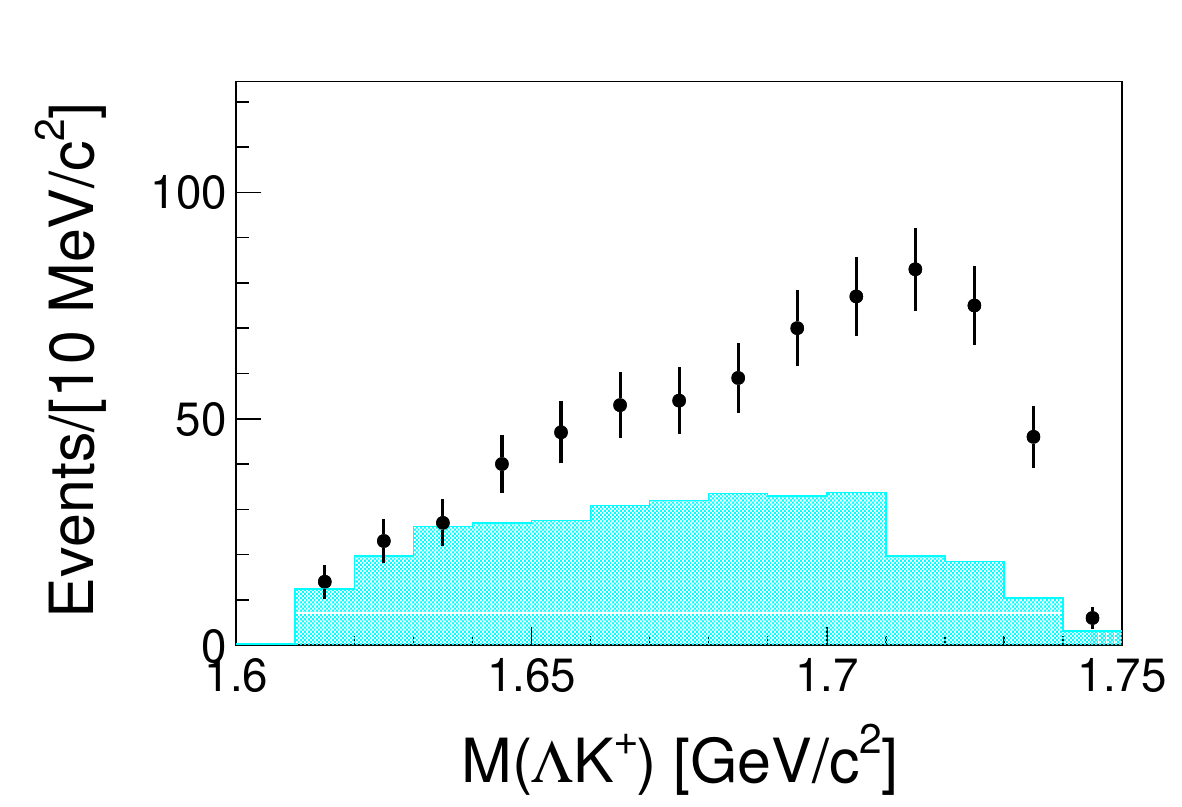} \put(-180,70){\bf Belle II} 
	\includegraphics[width=5.0cm]{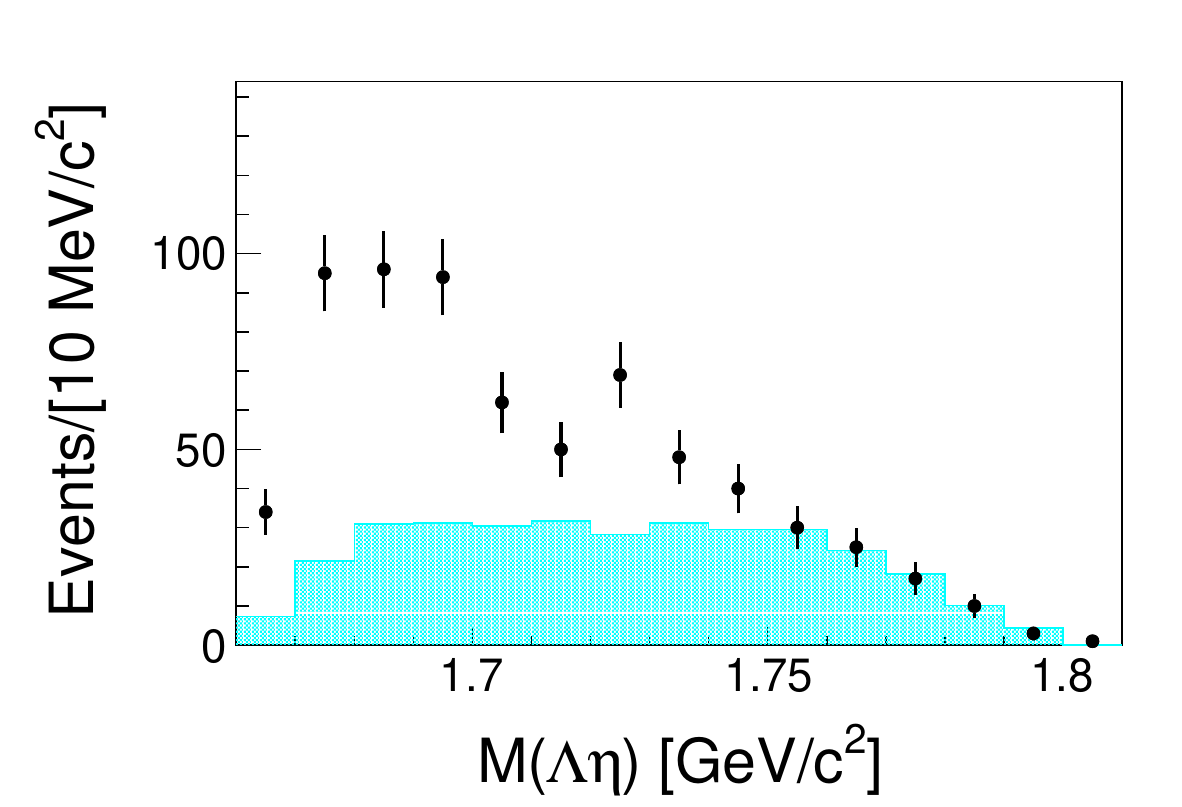}
	\includegraphics[width=5.0cm]{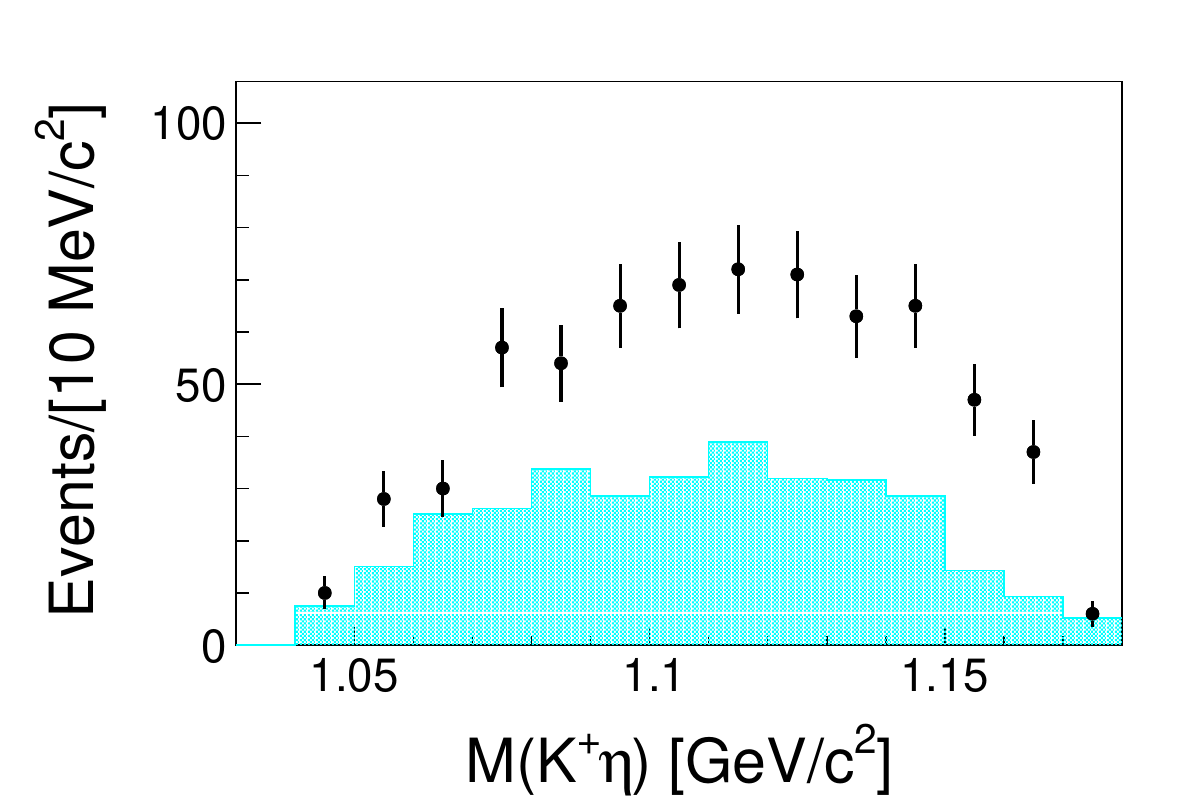}
	\caption{Invariant-mass distributions of the two-body combinations in (a) $\Lambda_c^+\to\Lambda K^+\pi^0$ and (b) $\Lambda_c^+\to\Lambda K^+\eta$ candidates for the Belle and Belle~II data samples. Black points with statistical error bars show the data distributions, while the cyan histograms show the normalized background distributions obtained from the inclusive simulation samples, as described in the text.
    The vertical black lines indicate the veto region of two-body decay $\Lambda_c^+\to\Xi^0K^+$, followed by $\Xi^0\to\Lambda\pi^0$. }
	\label{fig:data:subM}
\end{figure*}

Figure~\ref{fig:data:subM} shows the invariant-mass distributions of all pairwise combinations in the $\Lambda_c^+\to\Lambda K^+\pi^0$ and $\Lambda_c^+\to\Lambda K^+\eta$ candidates for the Belle and Belle~II data samples. 
These distributions exhibit significant deviations from a pure phase-space expectation.
Therefore, the Dalitz distributions observed in data are used to account for the observed non-phase-space structure in the reconstruction efficiency determination.

The reconstruction efficiencies of the reference mode $\Lambda_c^+ \to \Lambda \pi^+\pi^0 $ and two signal modes $\Lambda_c^+ \to \Lambda K^+\pi^0 $ and $\Lambda_c^+\to\Lambda K^+\eta$ are evaluated using simulated signal samples reweighted in two steps.
First, the simulated samples are reweighted according to the efficiency-corrected scaled momentum distribution measured with the control sample $\Lambda_c^+\to\Lambda\pi^+$.
The samples are then reweighted according to the background-subtracted Dalitz distribution observed in data within the signal region, 
where the Dalitz plot is constructed from the squared invariant masses of the  $\Lambda \pi^+/K^+$ and $\Lambda \pi^0/\eta$ combinations.
The resulting efficiency is calculated as
\begin{equation}
    \varepsilon=\frac{1}{\sum_{i}^{N_{\rm bins}}w_i/\varepsilon_i}.
\end{equation}
Here, $w_i$ is the weight in the $i$th Dalitz bin, obtained from the background-subtracted data as
\begin{equation}
w_i = (N_{{\rm data},i}-f_s N_{{\rm inclusive},i})/(N_{\rm data}-f_sN_{\rm inclusive}),
\end{equation}
where $N_{\rm data}$ and $N_{\rm inclusive}$ denote the numbers of events in the signal region for data and inclusive simulated samples, respectively, and $f_s$ is the global background-normalization factor. The quantity $\varepsilon_i$ is the reconstruction efficiency in the $i$th Dalitz bin obtained from the signal simulation. 
The distributions of normalized background in simulations are also shown in Fig.~\ref{fig:data:subM}.
Since the phase-space regions and distributions differ significantly between the events in the signal and sideband regions, the inclusive simulated samples are used for the background subtraction instead of the sideband region of the experimental data. The agreement between the background distributions in data and simulation is verified  using events in the sideband region and is found to be consistent within statistical uncertainties.

The reconstruction efficiencies are further corrected by multiplicative data-to-simulation correction factors.
We apply corrections associated with the PID requirements
obtained from the calibration channel $D^{*+}\to D^0(\to K^-\pi^+)\pi^+$.
The $\pi^0$ reconstruction efficiency is determined using the $\tau\to h\pi^0\nu_\tau$ decays in both Belle and Belle II, while the $\eta$ reconstruction efficiency is estimated from the studies of photon reconstruction efficiency using $e^+ e^- \to e^+ e^- \gamma $ events in Belle  and $e^+ e^- \to \mu^+ \mu^- \gamma $ events in Belle~II.
The correction factors for the $\pi^0$ veto are determined from fits to the $\pi^0$ and $\eta$ invariant-mass spectra. The overall correction factors in Belle are $(93.4\pm2.4)\%$, $(92.6\pm2.5)\%$, and $(92.7\pm4.3)\%$ for the reference channel $\Lambda_c^+\to\Lambda \pi^+\pi^0$ and the signal modes $\Lambda_c^+\to\Lambda K^+\pi^0$ and $\Lambda_c^+\to\Lambda K^+ \eta$, respectively. The corresponding correction factors in Belle II are 
$(90.9\pm2.4)\%$, $(91.4\pm2.7)\%$, and $(85.6\pm3.1)\%$. 
The difference in correction factor between Belle and Belle~II  is larger for the $\eta$ mode than for the other modes, mainly due to the correction factors associated with the $\pi^0$ veto, which are affected by the imperfect simulation of the larger photon background in Belle~II.
The quoted uncertainties are obtained from the corresponding control samples.

The yields of the signal decays, $\Lambda_c^+\to \Lambda K^+\pi^0$ and $\Lambda_c^+\to \Lambda K^+\eta$, and the reference decay, $\Lambda_c^+\to\Lambda\pi^+\pi^0$, are extracted from unbinned extended maximum-likelihood fits to the corresponding invariant-mass distributions. 
The probability density functions (PDFs) comprise a signal component and a combinatorial background for all channels, with additional feed-down and misidentification backgrounds included where applicable. 
A double bifurcated Gaussian function is used for the reference channel, $\Lambda_c^+\to\Lambda \pi^+\pi^0$. A bifurcated Gaussian combined with an additional Gaussian function is used for the $\Lambda_c^+\to\Lambda K^+\pi^0$ channel, while a sum of three Gaussian functions is used for the $\Lambda_c^+\to\Lambda K^+\eta$. In each case, the Gaussian components share a common mean.
All shape parameters, except for the mean, are fixed to the values obtained from the corresponding signal simulations.
To account for differences in the mass resolution between data and simulation, all width parameters fixed to the values obtained from simulation are multiplied by a common scale factor determined from a fit to the reference channel $\Lambda_c^+\to\Lambda\pi^+\pi^0$. 
The scale factor is subsequently fixed in the fits to the $\Lambda_c^+\to\Lambda K^+\pi^0$ and $\Lambda_c^+\to\Lambda K^+\eta$ channels.
The feed-down and misidentification background PDFs are determined from the corresponding exclusive simulated samples directly, with their peak positions treated as free parameters in the fit.
The feed-down component is included in the fits to the $\Lambda_c^+\to\Lambda\pi^+\pi^0$ and $\Lambda_c^+\to\Lambda K^+\pi^0$ channels, while the misidentification component is included only in the $\Lambda_c^+\to\Lambda K^+\pi^0$ fit. The yields of both background components are free parameters in the fit.
The combinatorial background is modeled with a third-order polynomial for the $\Lambda_c^+\to\Lambda\pi^+\pi^0$ channel and a first-order polynomial for the $\Lambda_c^+\to\Lambda K^+\pi^0$ and $\Lambda_c^+\to\Lambda K^+\eta$ channels, with all polynomial parameters and yields allowed to float.

The reference channel is fitted independently in the Belle and Belle~II data samples. 
The efficiency-corrected yields per unit integrated luminosity are found to be consistent, within uncertainties, between the two experiments using the data collected at the $\Upsilon(4S)$ energy. 

For the signal channels, simultaneous fits are performed to the Belle and Belle~II data samples. The ratio of the signal yields in the two samples is fixed 
to the expectation based on the fitted reference-mode yields and the relative reconstruction efficiencies of the signal and reference channels.
The signal yields obtained from the simultaneous fits are verified to be consistent with the sum of those extracted from independent fits to the Belle and Belle~II data.
The fitting procedure is validated using simulated samples, and no significant fit bias is observed.

The fit results are shown in Fig.~\ref{fig:Mlambdac:data}, together with the pull distributions, defined as $(N_{\rm data}-N_{\rm fit})/\sigma_{\rm data}$. Here, $N_{\rm data}$, $N_{\rm fit}$, and $\sigma_{\rm data}$ denote the number of data events, the fitted yield, and the uncertainty on $N_{\rm data}$, respectively, in each displayed bin. The fitted yields and reconstruction efficiencies are summarized in Table~\ref{tab:yield}.
The statistical significances are determined from the change in the likelihood between fits performed with and without the signal component.
A clear $\Lambda_c^+\to\Lambda K^+\pi^0$ signal is observed with a statistical significance well above 10 standard deviations ($10\sigma$).
We also report the first observation of the $\Lambda_c^+\to\Lambda K^+\eta$ decay, with a statistical significance exceeding $10\sigma$.

\begin{table*}[]
	\caption{
    Yields and reconstruction efficiencies for the reference $\Lambda_c^+\to\Lambda\pi^+\pi^0$ and signal decays $\Lambda_c^+\to\Lambda K^+\pi^0$ and $\Lambda_c^+\to\Lambda K^+\eta$  in the Belle and Belle~II data. 
    The signal mode yields
    are extracted from the simultaneous fit to the Belle and Belle~II data. 
    The quoted uncertainties are statistical only.
	}
	\centering
    \setlength{\tabcolsep}{14pt}
	\label{tab:yield}
	\begin{tabular}{lcccc}
		\hline\hline
		Channel & \multicolumn{2}{c}{Yield} & \multicolumn{2}{c}{Reconstruction efficiency (\%)} \\
		 & Belle & Belle II & Belle & Belle II\\
		 \hline
		 $\Lambda_c^+\to\Lambda\pi^+\pi^0$ & $(281.0\pm1.3)\times10^3$ & $(156.5\pm0.8)\times10^3$ & $5.38\pm0.02$ & $5.71\pm0.02$\\
		 $\Lambda_c^+\to\Lambda K^+\pi^0$ & \multicolumn{2}{c}{$5811\pm233$} & $3.77\pm0.02$ & $4.46\pm0.02$\\
		 $\Lambda_c^+\to\Lambda K^+\eta$  & \multicolumn{2}{c}{$1107\pm45$}   & $4.08\pm0.02$ & $3.85\pm0.02$ \\
		\hline\hline
	\end{tabular}
\end{table*}

\section{Branching Fraction Measurements}

The relative branching fractions of $\Lambda_c^+\to\Lambda K^+\pi^0$ and $\Lambda_c^+\to\Lambda K^+\eta$ , normalized to the reference decay $\Lambda_c^+\to\Lambda\pi^+\pi^0$, are calculated as 
\begin{flalign}
\label{eq:br}
&\frac{\mathcal{B}(\Lambda_c^+\to\Lambda K^+\pi^0)}{\mathcal{B}(\Lambda_c^+\to\Lambda\pi^+\pi^0)} = 
\frac{N_{K\pi}}{(N_{\pi\pi,1}\times \frac{\varepsilon_{K\pi,1}}{\varepsilon_{\pi\pi,1}} + N_{\pi\pi,2}\times \frac{\varepsilon_{K\pi,2}}{\varepsilon_{\pi\pi,2}})}, \\
\label{eq:br2}
&\frac{\mathcal{B}(\Lambda_c^+\to\Lambda K^+\eta)}{\mathcal{B}(\Lambda_c^+\to\Lambda\pi^+\pi^0)} =  &\nonumber\\ 
&\frac{N_{K\eta}}{(N_{\pi\pi,1}\times \frac{\varepsilon_{K\eta,1}}{\varepsilon_{\pi\pi,1}} + N_{\pi\pi,2}\times \frac{\varepsilon_{K\eta,2}}{\varepsilon_{\pi\pi,2}})}\times\frac{\mathcal{B}(\pi^0\to\gamma\gamma)}{\mathcal{B}(\eta\to\gamma\gamma)}.&
\end{flalign}
Here $N$ and $\varepsilon$ denote the fitted yields and reconstruction efficiencies, respectively. The subscripts $\pi\pi$, $K\pi$, and $K\eta$ correspond to the $\Lambda_c^+\to\Lambda\pi^+\pi^0$, $\Lambda_c^+\to\Lambda K^+\pi^0$, and $\Lambda_c^+\to\Lambda K^+\eta$ channels, respectively, while the indices $1$ and $2$ refer to the Belle and Belle~II data samples. The branching fractions of $\pi^0\to\gamma\gamma$ and $\eta\to\gamma\gamma$ are taken from the Particle Data Group (PDG)~\cite{pdg}. The branching fraction of $\Lambda\to p\pi^-$ cancels in the ratio and is therefore omitted.

\section{Systematic Uncertainties\label{sec:syst}}

Table~\ref{tab:syst} summarizes the systematic uncertainties on the ratio of branching fractions $\Lambda_c^+\to\Lambda K^+\pi^0 / \Lambda_c^+\to\Lambda \pi^+\pi^0 $ and $\Lambda_c^+\to\Lambda K^+\eta / \Lambda_c^+\to\Lambda \pi^+\pi^0$. 
The sources of these uncertainties include the limited size of the simulated samples; the Dalitz-plot correction; the data-to-simulation correction factors for PID, $\pi^0$ and $\eta$ reconstruction;  the $\pi^0$ veto requirement; the fitting procedure used to extract the signal and reference yields; the branching fractions of the intermediate-state decays ($\mathcal{B}_{\rm inter}$).
The uncertainties associated with the branching fractions of intermediate states and the fit procedure are common to the Belle and Belle~II measurements. All other systematic uncertainties are evaluated independently for the two experiments and then combined in quadrature with appropriate weights according to Eqs.~(\ref{eq:br},~\ref{eq:br2}).
The total systematic uncertainty is obtained by summing the individual contributions in quadrature.

\begin{table}
	\caption{The relative systematic and statistical uncertainties (\%) on the ratio of branching fractions. 
	}
	\centering
	\label{tab:syst}
	\begin{tabular}{l|cc}
		\hline\hline
		Source & $\frac{\mathcal{B}(\Lambda_c^+\to\Lambda K^+\pi^0)}{\mathcal{B}(\Lambda_c^+\to\Lambda\pi^+\pi^0)} $ & $\frac{\mathcal{B}(\Lambda_c^+\to\Lambda K^+\eta)}{\mathcal{B}(\Lambda_c^+\to\Lambda\pi^+\pi^0)} $ \\
		\hline
		Simulation sample size & 0.3 & 0.3 \\
        Dalitz correction & 1.9 & 1.1 \\
		PID 	& 1.2	& 1.2	\\
        $\pi^0$ and $\eta$ reconstruction & 0.6	& 4.4	\\
		$\pi^0$ veto & - & 1.1	\\
   		Fit range & 0.1 & 0.6 \\
		Background shape & 2.8 & 0.1 \\
		Mass resolution & 0.2 & 0.1\\
		Reference mode fitting & 0.8 & 0.8 \\
		$\mathcal{B}_{\rm inter}$ & - & 0.5 \\
		\hline
		Total & 3.7 & 5.0 \\
        \hline
        Statistical & 4.0 & 4.0\\
		\hline\hline
	\end{tabular}
\end{table}

The systematic uncertainty due to the limited size of the simulated samples is evaluated using the binomial uncertainty on the reconstruction efficiency.

The systematic uncertainty associated with the efficiency correction over the Dalitz plane is estimated by varying the binning scheme from the nominal $20\times20$ grid to $15\times15$ and $25\times25$, and taking the resulting change in the reconstruction efficiency as the systematic uncertainty. 

The effect of the background normalization used in the Dalitz-plot correction is evaluated by recalculating the reconstruction efficiency using bin-dependent normalization factors obtained with a coarser $3\times3$ binning. The resulting variation is below 0.2\%, which is within the uncertainty arising from the limited simulation sample size, and is therefore neglected.

Owing to the relative nature of the measurement, most corrections associated with the detection efficiencies and selection requirements, as well as their associated  uncertainties, cancel. 
The uncertainties associated with the utilized correction factors are evaluated separately for the signal and reference channels. 
The corresponding uncertainties that are estimated using the same control sample in the signal and reference-channel efficiencies are assumed to be fully correlated and are propagated to the efficiency ratio by adding their contributions linearly.

Systematic uncertainties associated with PID are estimated to be 0.8\% (0.1\%) for each pion  and 0.9\% (1.3\%) for each kaon  in Belle (Belle~II)~\cite{b1PID:Nakano:2002jw,b2hid:Belle-II:2025tpe}, while the correction and associated uncertainty for proton PID cancel. 
The resulting systematic uncertainty associated with the PID efficiency correction is 1.2\% for both $\Lambda_c^+\to\Lambda K^+\pi^0$  and  $\Lambda_c^+\to\Lambda K^+\eta $  decays.

The uncertainty on the $\pi^0$ reconstruction efficiency is estimated to be 2.2--2.4\%. 
A conservative uncertainty of 0.8\% is assigned to the $\pi^0$ reconstruction in $\varepsilon_{K\pi}/\varepsilon_{\pi\pi}$ for both Belle and Belle~II to account for the slight difference in the $\pi^0$ momentum distributions between the signal and reference channels. 
The uncertainty on the $\eta$ reconstruction efficiency is derived from the photon reconstruction uncertainty and amounts to 2.0\% and 0.7\% per photon in Belle and Belle~II, respectively.
The combined uncertainties due to the $\pi^0$ and $\eta$ reconstruction are assigned as 0.6\% and 4.4\%, respectively, for $\Lambda_c^+\to\Lambda K^+\pi^0$  and  $\Lambda_c^+\to\Lambda K^+\eta $  measurements.

The correction factor associated with the $\pi^0$ veto cancels in the efficiency ratio $\varepsilon_{K\pi}/\varepsilon_{\pi\pi}$. For $\varepsilon_{K\eta}/\varepsilon_{\pi\pi}$, the corresponding uncertainty is determined from fits to the $\pi^0$ and $\eta$ invariant-mass spectra, yielding a total uncertainty of 1.1\%.

The systematic uncertainty associated with charged-track reconstruction largely cancels because all channels contain three charged tracks with similar kinematic distributions.
The uncertainty associated with $\Lambda$ reconstruction is evaluated following the method of Ref.~\cite{Belle:2025qrq}, and is found to be below 0.1\%, and therefore neglected.
The $\Xi^0\to\Lambda\pi^0$ contribution occupies only a small region of the three-body phase space, and its contribution to the fitted $\Lambda_c^+$ yield is below 0.1\%. The corresponding uncertainty is therefore neglected.

The systematic uncertainty associated with the fit procedure is evaluated by comparing the nominal fit with several alternative configurations. 
The uncertainty due to the background shape of the feed-down contribution is evaluated by increasing the polynomial order used to describe the combinatorial background by one.
The uncertainty associated with the fit range is evaluated by varying the fit range by $\pm20$\mevcc. The uncertainty due to the mass resolution is estimated by varying the resolution scale factor within its uncertainty. The resulting changes in the fitted yields are assigned as systematic uncertainties, and the total fit uncertainty is obtained by summing the individual contributions in quadrature.

The uncertainties on the branching fractions of intermediate-state decays are taken from the PDG~\cite{pdg}. The uncertainty on $\mathcal{B}(\Lambda_c^+\to\Lambda\pi^+\pi^0)$ (4.8\%) is treated separately and is included only in the measurement of the absolute branching fractions.

\section{Results}

The measured relative branching fractions are 
\begin{equation}
\frac{\mathcal{B}(\Lambda_c^+\to\Lambda K^+\pi^0)}{\mathcal{B}(\Lambda_c^+\to\Lambda\pi^+\pi^0)}=(1.82 \pm 0.07 \pm 0.07)\%
\end{equation}
and
\begin{equation}
\frac{\mathcal{B}(\Lambda_c^+\to\Lambda K^+\eta)}{\mathcal{B}(\Lambda_c^+\to\Lambda\pi^+\pi^0)}=(8.73 \pm 0.35 \pm 0.43)\times10^{-3},
\end{equation}
where the first uncertainty is statistical and the second is systematic. 

Using the world-average value of $\mathcal{B}(\Lambda_c^+\to\Lambda\pi^+\pi^0)$ from the PDG~\cite{pdg}, we determine the absolute branching fractions
\begin{equation}
\mathcal{B}(\Lambda_c^+\to\Lambda K^+\pi^0)=(1.29\pm0.05\pm0.05\pm0.06)\times10^{-3}
\end{equation}
and
\begin{equation}
\mathcal{B}(\Lambda_c^+\to\Lambda K^+\eta)=(6.20\pm0.25\pm0.31\pm0.30)\times10^{-4},
\end{equation}
where the first uncertainty is statistical, the second is systematic, and the third arises from the uncertainty on $\mathcal{B}(\Lambda_c^+\to\Lambda\pi^+\pi^0)$.

\begin{figure}[]	
	\centering
	\raggedright
	\includegraphics[width=6.0cm]{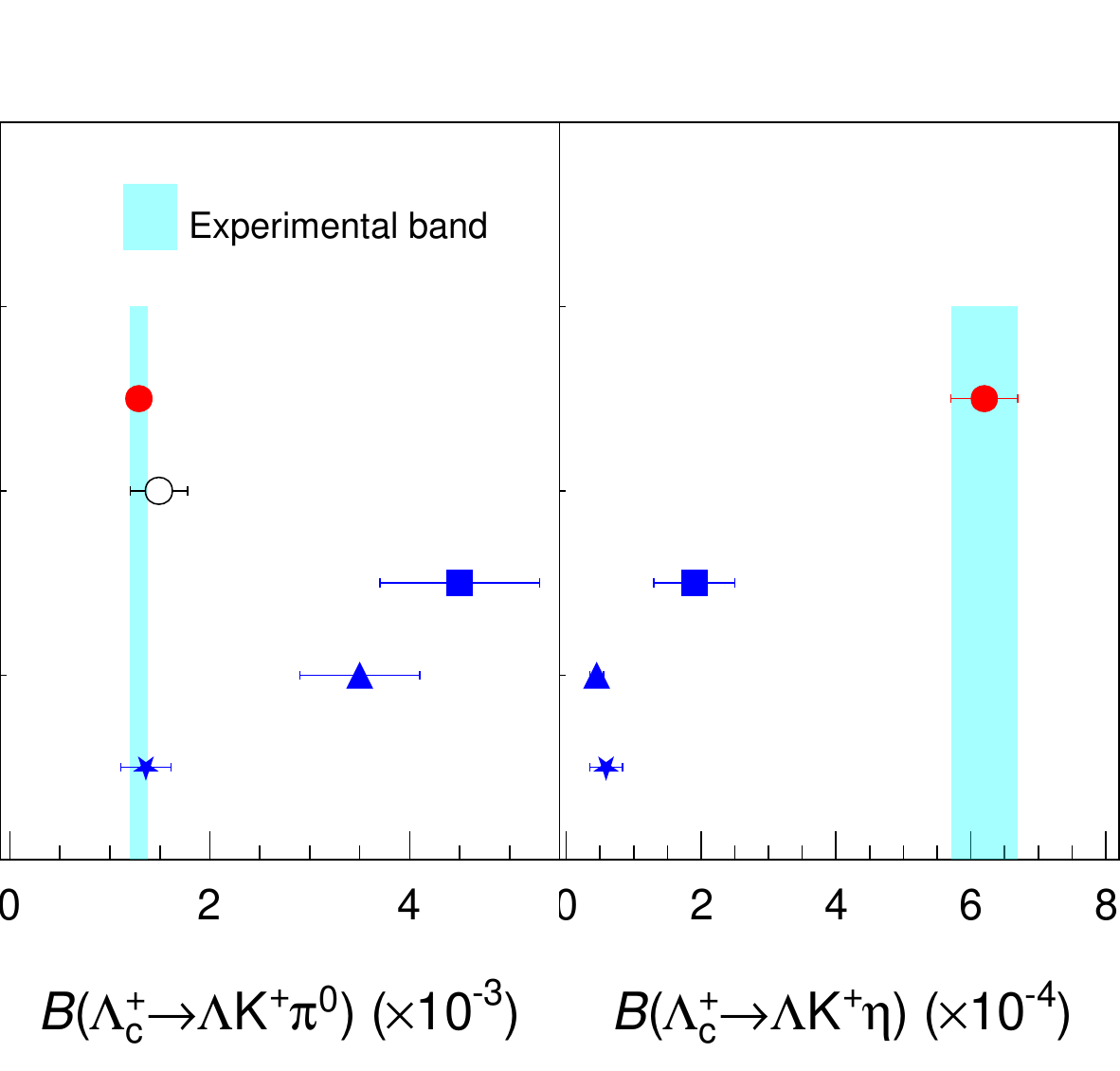}	
	\put(-170,135){\bf (a)}
	\put(-85,135){\bf (b)} 
	\put(5,100){This work}  
	\put(5,85){BESIII~\cite{Exp:Lc2lbdkpi:BESIII:2023sdr} }  
	\put(5,70){Geng 2019~\cite{Theory:Predict1:Geng:2018upx}} 
	\put(5,55){Cen 2019~\cite{Thoery:Predict2:Cen:2019ims}} 
	\put(5,40){Geng 2024~\cite{Theory:Predict3:Geng:2024sgq}} 
	\put(-170,150){{\bf Belle + Belle II} $\rm \bf \int Ldt=1.55$\invab}
	\caption{Comparison of the branching fractions for (a) $\Lambda_c^+\to\Lambda K^+\pi^0$  and (b) $\Lambda_c^+\to\Lambda K^+\eta$ between this measurement, the previous BESIII result~\cite{Exp:Lc2lbdkpi:BESIII:2023sdr} and theoretical predictions based on the $SU(3)_F$ flavor-symmetry approach in Refs.~\cite{Theory:Predict1:Geng:2018upx,Thoery:Predict2:Cen:2019ims,Theory:Predict3:Geng:2024sgq}, shown as red dots, a hollow circle, and blue squares, triangles, and stars, respectively. 
    The cyan box indicates the $\pm1\sigma$ experimental band measured in this analysis.
    The uncertainties on the experimental measurements include the statistical, systematic, and $\mathcal{B}(\Lambda_c^+\to\Lambda \pi^+\pi^0)$ uncertainties combined in quadrature.
	}
	\label{fig:result}
\end{figure}

Figure~\ref{fig:result} compares the measured branching fractions of $\Lambda_c^+\to\Lambda K^+\pi^0$ and $\Lambda_c^+\to\Lambda K^+\eta$ with the previous BESIII result~\cite{Exp:Lc2lbdkpi:BESIII:2023sdr} and theoretical predictions based on the $SU(3)_F$ flavor symmetry approach~\cite{Theory:Predict1:Geng:2018upx,Thoery:Predict2:Cen:2019ims,Theory:Predict3:Geng:2024sgq}. 
The measured branching fraction of $\Lambda_c^+\to\Lambda K^+\pi^0$ is consistent with the previous BESIII measurement and the latest theoretical prediction~\cite{Theory:Predict3:Geng:2024sgq}, while significantly lower than the earlier predictions~\cite{Theory:Predict1:Geng:2018upx,Thoery:Predict2:Cen:2019ims}. In contrast, the measured branching fraction of $\Lambda_c^+\to\Lambda K^+\eta$ is significantly larger than all current theoretical predictions~\cite{Theory:Predict1:Geng:2018upx,Thoery:Predict2:Cen:2019ims,Theory:Predict3:Geng:2024sgq} by $5.5\sigma$, $11.2\sigma$, and $10.1\sigma$, respectively, motivating further theoretical studies of charmed-baryon three-body decays.
The observed deviations in the invariant-mass spectra of the sub-combinations from the phase-space simulation may also encourage future experimental and theoretical studies aimed at understanding the underlying decay dynamics, including possible intermediate resonant structures.

\section{Conclusion}

We measure the branching fractions of the three-body singly Cabibbo-suppressed decays $\Lambda_c^+\to\Lambda K^+\pi^0$ and $\Lambda_c^+\to\Lambda K^+\eta$ using the combined Belle and Belle~II data sample, corresponding to an integrated luminosity of 1.55\invab.

The branching fraction ratios relative to the Cabibbo-favored decay $\Lambda_c^+\to\Lambda\pi^+\pi^0$ are measured to be
${\mathcal{B}(\Lambda_c^+\to\Lambda K^+\pi^0)}/{\mathcal{B}(\Lambda_c^+\to\Lambda\pi^+\pi^0)} = (1.82 \pm 0.07 \pm 0.07)\%$ and ${\mathcal{B}(\Lambda_c^+\to\Lambda K^+\eta)}/{\mathcal{B}(\Lambda_c^+\to\Lambda\pi^+\pi^0)} = (8.73 \pm 0.35 \pm 0.43)\times10^{-3}$.

Using the world-average value of $\mathcal{B}(\Lambda_c^+\to\Lambda\pi^+\pi^0)$~\cite{pdg}, we obtain the absolute branching fractions
$\mathcal{B}(\Lambda_c^+ \to \Lambda K^+ \pi^0)=(1.29\pm0.05\pm0.05\pm0.06)\times10^{-3}$
and
$\mathcal{B}(\Lambda_c^+ \to \Lambda K^+ \eta)=(6.20\pm0.25\pm0.31\pm0.30)\times10^{-4}$.
The first, second, and third uncertainties are statistical, systematic, and due to the uncertainty on $\mathcal{B}(\Lambda_c^+\to\Lambda\pi^+\pi^0)$, respectively. 
The measured branching fraction of $\Lambda_c^+\to\Lambda K^+\pi^0$ is consistent with the previous BESIII result~\cite{Exp:Lc2lbdkpi:BESIII:2023sdr} while significantly improving its precision. The $\Lambda_c^+\to\Lambda K^+\eta$ decay is observed for the first time with a statistical significance larger than $10\sigma$.

\begin{acknowledgments}
This work, based on data collected using the Belle II detector, which was built and commissioned prior to March 2019,
and data collected using the Belle detector, which was operated until June 2010,
was supported by
Higher Education and Science Committee of the Republic of Armenia Grant No.~23LCG-1C011;
Australian Research Council and Research Grants
No.~DP200101792, 
No.~DP210101900, 
No.~DP210102831, 
No.~DE220100462, 
No.~LE210100098, 
and
No.~LE230100085; 
Austrian Federal Ministry of Education, Science and Research,
Austrian Science Fund (FWF) Grants
DOI:~10.55776/P34529,
DOI:~10.55776/J4731,
DOI:~10.55776/J4625,
DOI:~10.55776/M3153,
and
DOI:~10.55776/PAT1836324,
and
Horizon 2020 ERC Starting Grant No.~947006 ``InterLeptons'';
Natural Sciences and Engineering Research Council of Canada, Digital Research Alliance of Canada, and Canada Foundation for Innovation;
National Key R\&D Program of China under Contract No.~2024YFA1610503,
and
No.~2024YFA1610504
National Natural Science Foundation of China and Research Grants
No.~11575017,
No.~11761141009,
No.~11705209,
No.~11975076,
No.~12135005,
No.~12150004,
No.~12161141008,
No.~12405099,
No.~12475093,
and
No.~12175041,
and Shandong Provincial Natural Science Foundation Project~ZR2022JQ02;
the Czech Science Foundation Grant No. 22-18469S,  Regional funds of EU/MEYS: OPJAK
FORTE CZ.02.01.01/00/22\_008/0004632 
and
Charles University Grant Agency project No. 246122;
European Research Council, Seventh Framework PIEF-GA-2013-622527,
Horizon 2020 ERC-Advanced Grants No.~267104 and No.~884719,
Horizon 2020 ERC-Consolidator Grant No.~819127,
Horizon 2020 Marie Sklodowska-Curie Grant Agreement No.~700525 ``NIOBE''
and
No.~101026516,
and
Horizon 2020 Marie Sklodowska-Curie RISE project JENNIFER2 Grant Agreement No.~822070 (European grants);
L'Institut National de Physique Nucl\'{e}aire et de Physique des Particules (IN2P3) du CNRS
and
L'Agence Nationale de la Recherche (ANR) under Grant No.~ANR-21-CE31-0009 (France);
BMFTR, DFG, HGF, MPG, and AvH Foundation (Germany);
Department of Atomic Energy under Project Identification No.~RTI 4002,
Department of Science and Technology,
and
UPES SEED funding programs
No.~UPES/R\&D-SEED-INFRA/17052023/01 and
No.~UPES/R\&D-SOE/20062022/06 (India);
Israel Science Foundation Grant No.~2476/17,
U.S.-Israel Binational Science Foundation Grant No.~2016113, and
Israel Ministry of Science Grant No.~3-16543;
Istituto Nazionale di Fisica Nucleare and the Research Grants BELLE2,
and
the ICSC – Centro Nazionale di Ricerca in High Performance Computing, Big Data and Quantum Computing, funded by European Union – NextGenerationEU;
Japan Society for the Promotion of Science, Grant-in-Aid for Scientific Research Grants
No.~16H03968,
No.~16H03993,
No.~16H06492,
No.~16K05323,
No.~17H01133,
No.~17H05405,
No.~18K03621,
No.~18H03710,
No.~18H05226,
No.~19H00682, 
No.~20H05850,
No.~20H05858,
No.~22H00144,
No.~22K14056,
No.~22K21347,
No.~23H05433,
No.~26220706,
and
No.~26400255,
and
the Ministry of Education, Culture, Sports, Science, and Technology (MEXT) of Japan;  
National Research Foundation (NRF) of Korea Grants
No.~2021R1-F1A-1064008, 
No.~2022R1-A2C-1003993,
No.~2022R1-A2C-1092335,
No.~RS-2016-NR017151,
No.~RS-2018-NR031074,
No.~RS-2021-NR060129,
No.~RS-2023-00208693,
No.~RS-2024-00354342
and
No.~RS-2025-02219521,
Radiation Science Research Institute,
Foreign Large-Size Research Facility Application Supporting project,
the Global Science Experimental Data Hub Center, the Korea Institute of Science and
Technology Information (K25L2M2C3 ) 
and
KREONET/GLORIAD;
Universiti Malaya RU grant, Akademi Sains Malaysia, and Ministry of Education Malaysia;
Frontiers of Science Program Contracts
No.~FOINS-296,
No.~CB-221329,
No.~CB-236394,
No.~CB-254409,
and
No.~CB-180023, and SEP-CINVESTAV Research Grant No.~237 (Mexico);
the Polish Ministry of Science and Higher Education and the National Science Center;
the Ministry of Science and Higher Education of the Russian Federation
and
the HSE University Basic Research Program, Moscow;
University of Tabuk Research Grants
No.~S-0256-1438 and No.~S-0280-1439 (Saudi Arabia), and
Researchers Supporting Project number (RSPD2025R873), King Saud University, Riyadh,
Saudi Arabia;
Slovenian Research Agency and Research Grants
No.~J1-50010
and
No.~P1-0135;
Ikerbasque, Basque Foundation for Science,
State Agency for Research of the Spanish Ministry of Science and Innovation through Grant No. PID2022-136510NB-C33, Spain,
Agencia Estatal de Investigacion, Spain
Grant No.~RYC2020-029875-I
and
Generalitat Valenciana, Spain
Grant No.~CIDEGENT/2018/020;
the Swiss National Science Foundation;
The Knut and Alice Wallenberg Foundation (Sweden), Contracts No.~2021.0174 and No.~2021.0299;
National Science and Technology Council,
and
Ministry of Education (Taiwan);
Thailand Center of Excellence in Physics;
TUBITAK ULAKBIM (Turkey);
National Research Foundation of Ukraine, Project No.~2020.02/0257,
and
Ministry of Education and Science of Ukraine;
the U.S. National Science Foundation and Research Grants
No.~PHY-1913789 
and
No.~PHY-2111604, 
and the U.S. Department of Energy and Research Awards
No.~DE-AC06-76RLO1830, 
No.~DE-SC0007983, 
No.~DE-SC0009824, 
No.~DE-SC0009973, 
No.~DE-SC0010007, 
No.~DE-SC0010073, 
No.~DE-SC0010118, 
No.~DE-SC0010504, 
No.~DE-SC0011784, 
No.~DE-SC0012704, 
No.~DE-SC0019230, 
No.~DE-SC0021274, 
No.~DE-SC0021616, 
No.~DE-SC0022350, 
No.~DE-SC0023470; 
and
the Vietnam Academy of Science and Technology (VAST) under Grants
No.~NVCC.05.02/25-25
and
No.~DL0000.05/26-27.

These acknowledgements are not to be interpreted as an endorsement of any statement made
by any of our institutes, funding agencies, governments, or their representatives.

We thank the SuperKEKB team for delivering high-luminosity collisions;
the KEK cryogenics group for the efficient operation of the detector solenoid magnet and IBBelle on site;
the KEK Computer Research Center for on-site computing support; the NII for SINET6 network support;
and the raw-data centers hosted by BNL, DESY, GridKa, IN2P3, INFN, 
PNNL/EMSL, 
and the University of Victoria.

\end{acknowledgments}

\bibliographystyle{apsrev4-2}
\bibliography{Ref}

\end{document}